\documentclass[aps,showkeys]{revtex4}

\usepackage[small]{caption}
\usepackage{amsmath,amsfonts,amscd,amssymb,epsf, epsfig,mathrsfs,graphicx}

\newcommand{\pa}{\partial}

\begin{document}

\title{Electromagnetic Casimir piston in higher dimensional spacetimes}

\author{L. P. Teo}\email{ LeePeng.Teo@nottingham.edu.my}
\address{Department of Applied Mathematics, Faculty of Engineering, University of Nottingham Malaysia Campus, Jalan Broga, 43500, Semenyih, Selangor Darul Ehsan, Malysia.}

\begin{abstract}
We consider the Casimir effect of the electromagnetic field   in a higher dimensional spacetime of the form $M\times \mathcal{N}$, where $M$ is the $4$-dimensional Minkowski spacetime and $\mathcal{N}$ is an $n$-dimensional compact manifold.   The Casimir force acting on a planar piston that can move freely inside a closed cylinder with the same cross section is investigated. Different combinations of perfectly conducting boundary conditions and infinitely permeable boundary conditions are imposed on the cylinder and the piston. It is verified that if the piston and the cylinder have the same boundary conditions, the piston is always going to be pulled towards the closer end of the cylinder. However, if the piston and the cylinder have different boundary conditions, the piston is always going to be pushed to the middle of the cylinder. By taking the limit where one end of the cylinder tends to infinity, one obtains the Casimir force acting between two parallel plates inside an infinitely long cylinder. The asymptotic behavior of this Casimir force in the high temperature regime and the low temperature regime are investigated for the case where  the cross section of the cylinder in $M$ is large. It is found that if the separation between the plates is much smaller than the size of $\mathcal{N}$, the leading term of the Casimir force is the same as the Casimir force on a pair of large parallel plates in the $(4+n)$-dimensional Minkowski spacetime. However, if the size of $\mathcal{N}$ is much smaller than the separation between the plates, the leading term of the Casimir force is $1+h/2$ times the Casimir force on a pair of large parallel plates in the $4$-dimensional Minkowski spacetime, where $h$ is the first Betti number of $\mathcal{N}$. In the limit the manifold $\mathcal{N}$ vanishes, one does not obtain the Casimir force in the $4$-dimensional Minkowski spacetime if $h$ is nonzero.

\end{abstract}

 \maketitle

\section{Introduction}

In 1948, Casimir proposed the existence of a force of magnitude
$$F=\frac{\pi^2\hbar cA}{240a^4}$$between two parallel perfectly conducting plates of area $A$ which are separated by a distance $a$  due to the vacuum fluctuations of electromagnetic field \cite{14}. Since 1970s, Casimir effect has aroused the interest of many researchers for its close relations with many other areas of physics such as quantum field theory, atomic physics, condensed matter physics, nanotechnology, astrophysics and mathematical physics \cite{15}. Although the original proposal of Casimir considered only the electromagnetic field in the $4D$ Minkowski spacetime,  nowadays the scope of Casimir effect includes all other quantum fields in arbitrary spacetimes of arbitrary dimensions. However, most of the works on Casimir effect in higher dimensional spacetimes considered only scalar fields. In some works, the simple relation between the Casimir force on a pair of large parallel plates due to a massless scalar field and the Casimir force on a pair of large parallel plates due to an electromagnetic field is wrongly extended to other geometric configurations.

By definition, the zero temperature Casimir energy is a divergent sum of the zero point energies of a quantum field. There are various methods such as cut-off method and zeta regularization method to remove the divergence and extract a physically meaningful Casimir energy. However, these divergence removal procedures can sometimes lead to ambiguities. In 2004, the piston configuration was introduced \cite{16} and it quickly attracted a lot of attention because the divergence of the Casimir energy in this configuration can be unambiguously removed. The zero temperature Casimir force acting on a piston due to scalar fields or electromagnetic fields in the $4D$ Minkowski spacetime was soon investigated in \cite{17,18,19,20}. This was then extended to rectangular piston in Minkowski spacetimes of arbitrary dimensions \cite{21,22} and to the finite temperature effect \cite{13}. Lately, there is an interest in considering the piston configuration in spacetimes with extra dimensions such as the Kaluza-Klein spacetime and the Randall-Sundrum spacetime \cite{23,24,25,26,27,28,29,30}. However, the works in this direction were restricted to scalar fields. To the best of our knowledge, no work has considered the electromagnetic Casimir effect on a piston in higher dimensional spacetimes.  The electromagnetic Casimir effect on a pair of large parallel perfectly conducting plates in the Kaluza-Klein spacetime with internal space $S^1$ and in the Randall-Sundrum spacetime have been considered in \cite{31,32}. As pointed out in \cite{32}, for an electromagnetic field in spacetimes with extra dimensions, one can either treat the field as a bulk field and impose the perfectly conducting boundary conditions introduced in \cite{4}, or one can use dimensional reduction to reduce the electromagnetic field to a tower of massive vector fields in the $4D$ Minkowski spacetime and impose the $4D$ perfectly conducting conditions on the massive vector fields. These two approaches lead to different Casimir effects. The first approach is a genuine higher dimensional Casimir effect whereas the second approach is essentially the Casimir effect of $4D$ massive vector fields.

\begin{figure}[h]
\epsfxsize=0.5\linewidth \epsffile{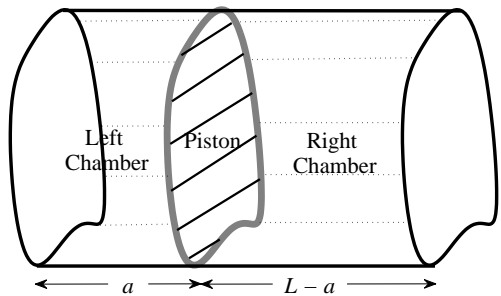} \caption{\label{f1} A piston system. }\end{figure}

In this article, we are interested in the Casimir effect in a higher dimensional spacetime due to the vacuum fluctuations of the electromagnetic field in a piston system.
The  spacetime $\mathcal{M}$ is assumed to have the form $M\times \mathcal{N}$, where $M$ is the $4D$ Minkowski spacetime and $\mathcal{N}$ is a manifold of dimension $n$, assumed to be compact and connected.
The piston system  consists of a cylinder of length $L$ and a piston  which can move freely inside the cylinder (See Fig. \ref{f1}). The position of the piston is given by $x^1=a$. The cross section of the cylinder and the piston  are the same and assumes the general form $\Omega\times\mathcal{N}$, where   $\Omega$ is  a two-dimensional simply connected  domain with boundary $\pa\Omega$ a smooth curve.
For the boundary conditions on the walls of the cylinder and the piston, we impose either the perfectly conducting boundary conditions or the infinitely permeable boundary conditions
proposed by Ambj{\o}rn and  Wolfram \cite{4}.

The  Casimir energy of the piston system is given by the sum of the Casimir  energy inside the left chamber, the Casimir energy inside the right chamber and the Casimir energy outside the cylinder. The latter does not depend on $a$ and therefore will not contribute to the Casimir force acting on the piston \cite{16}. Omitting this term, we have
\begin{equation}\label{eq1_24_1}
E_{\text{Cas}}^{\text{piston}}=E_{\text{Cas}}^{\text{cylinder}}(a)+E_{\text{Cas}}^{\text{cylinder}}(L-a),
\end{equation}where $E_{\text{Cas}}^{\text{cylinder}}(a)$ is the Casimir energy inside a cylinder of length $a$.
Using zeta regularization method, it is given by
\begin{equation}\label{eq1_24_2}
E_{\text{Cas}}^{\text{cylinder}}(a)=-\frac{T}{2}\left(\zeta_T'(0;a)+\log[\lambda^2]\zeta_T(0;a)\right),
\end{equation}where $\lambda$ is a normalization constant and $\zeta_T(s)$ is the zeta function
\begin{equation*}
\zeta_T(s;a)=\sum_{\omega}\sum_{p=-\infty}^{\infty}\left(\omega^{2}+[2\pi p T]^2\right)^{-s},
\end{equation*}which contains a sum over all $\omega$ that are the eigenfrequencies of the electromagnetic field inside the cylinder.

In Section \ref{BF}, we   discuss the gauge fixing of the electromagnetic in the spacetime $\mathcal{M}$. This enables us to write down the eigen-modes of the electromagnetic field in Section \ref{s3}. We then proceed to compute the Casimir energy and the Casimir force in Section \ref{s4}. A discussion of the asymptotic behavior of the Casimir force in different limits is also given in Section \ref{s4}. In Section \ref{s5}, we consider the specific examples where $\mathcal{N}$ is an $n$-torus or an $n$-sphere.

In the following, we use the units where $\hbar=c=k_B=1$.

\section{Gauge fixing of the electromagnetic field}\label{BF}
In this section, we will discuss the gauge fixing of the electromagnetic field in the spacetime $\mathcal{M}=M\times \mathcal{N}$, where $M$ is the $4D$ Minkowski spacetime and $\mathcal{N}$ is an $n$-dimensional compact connected manifold.

Let $$ds^2=g_{\mu\nu}dz^{\mu}dz^{\nu}=\eta_{\alpha\beta}dx^{\alpha}dx^{\beta} - G_{ab}dy^ady^b$$ be  the   metric on $ M\times \mathcal{N}$ with $\eta_{\alpha\beta}=\text{diag}(1, -1, -1, -1)$ the usual four-dimensional  metric on $M$;  and $ds_{\mathcal{N}}^2=G_{ab}dy^ady^b$ a Riemannian metric on $\mathcal{N}$.  $x$ denotes collectively the coordinates on $M$, $y$ denotes collectively the coordinates on $\mathcal{N}$ and $z=(x,y)$.
The action of the electromagnetic field is given  by
\begin{equation}\label{eq1_12_8}
S=-\frac{1}{4}\int  \sqrt{|g|}F_{\mu\nu}F^{\mu\nu} d^{N}z,
\end{equation} where $N=4+n$, $F_{\mu\nu}  = \pa_{\mu}A_{\nu}-\pa_{\nu}A_{\mu}  $ is an anti-symmetric rank two tensor and $F^{\mu\nu}=g^{\mu\kappa} g^{\nu\eta}F_{\kappa\eta}$.  The equation of motion is
\begin{equation}\label{eq1_12_7}
 \frac{1}{\sqrt{|g|}}\frac{\pa}{\pa z^{\mu}}\left( \sqrt{|g|}F^{\mu\nu}\right)   =0.
\end{equation}The perfectly conducting boundary condition is given by \cite{4}:
\begin{equation}\label{eq2_21_1}
\text{n}^{\mu}(*F)_{\mu \nu_1\ldots \nu_{N-3}}=0,
\end{equation}and the infinitely permeable boundary condition is given by \cite{4}:
\begin{equation}\label{eq2_21_2}
\text{n}^{\mu}F_{\mu\nu}=0.
\end{equation}Here $\text{n}^{\mu}$ is a unit  vector normal to the surface, and $*F$ is the dual tensor of $F$ defined in \eqref{eq1_25_3}.

Using the language of differential geometry \cite{1,2},   the electromagnetic field $F_{\mu\nu}dz^{\mu} dz^{\nu}  $ is an exterior two form on $\mathcal{M}$ and $A_{\mu}dz^{\mu}$ is an exterior one-form on $\mathcal{M}$. Moreover, $F_{\mu\nu}dz^{\mu}dz^{\nu}=d(A_{\mu}dz^{\mu})$.

Given an exterior $k$-form $\psi=\psi_{\mu_1\ldots \mu_k} dz^{\mu_1}\ldots dz^{\mu_k}$,   the dual $(N-k)$--form $*\psi=(*\psi)_{\nu_1\ldots\nu_{N-k}}dz^{\nu_1}\ldots dz^{\nu_{N-k}}$ is given by
\begin{equation}\label{eq1_25_3}
(*\psi)_{\nu_1\ldots\nu_{N-k}} = \frac{1}{k!}\sqrt{|g|} \varepsilon_{\nu_1\ldots\nu_{N-k}\mu_1\ldots \mu_k} \psi^{\mu_1\ldots\mu_k},
\end{equation}where $\varepsilon_{\mu_1\ldots\mu_{N}} $ is a totally antisymmetric rank-$N$ tensor which is equal to one if and only if $\mu_1, \ldots,\mu_{N}$ is an even permutation of $(0, 1, 2, \ldots,N-1)$. Using the $*$-operator and the exterior differentiation $d$, one can define the codifferential operator  $\delta$ by
\begin{equation*}
\delta\psi = (-1)^{N(k-1)}* d * \psi,
\end{equation*}which maps a $k$-form $\psi$ to a $k-1$-form. More explicitly,
\begin{equation*}
(\delta \psi)^{\mu_1\ldots\mu_{k-1}}= \frac{1}{\sqrt{|g|}}\frac{\pa}{\pa z^{\nu}}\left(\sqrt{|g|} \psi^{\nu\mu_1\ldots\mu_{k-1}}\right).
\end{equation*}The Laplacian operator $\Delta$ mapping $k$-forms to $k$-forms is then defined as
\begin{equation*}
\Delta=d\delta+\delta d.
\end{equation*}  Using these notations, the equation of motion \eqref{eq1_12_7} is equivalent to
\begin{equation}\label{eq1_12_9}
\delta F=\delta dA=0.
\end{equation} The action \eqref{eq1_12_8} is invariant under the gauge transformation $A \mapsto A+d\phi$ for any function $\phi$ on $ \mathcal{M}$. To fix the gauge, notice that for any one form $A$, it is always possible to find a function $\phi$ satisfying \begin{equation}\label{eq1_12_10}\Delta\phi=\delta d \phi = -\delta A.\end{equation} This implies that we can impose the Lorentz gauge \begin{equation}\label{eq1_12_11}\delta A=\frac{1}{\sqrt{|g|}}\pa_{\mu}\sqrt{|g|}A^{\mu}=0.\end{equation}The equation of motion  \eqref{eq1_12_9} is then equivalent to
 \begin{equation}\label{eq1_12_13}
 \Delta A = 0,
 \end{equation}i.e., $A^{\mu}dz^{\mu}$ is a harmonic one-form on $ \mathcal{M}$. However, \eqref{eq1_12_10} only defines $\phi$ up to those solutions satisfying $\Delta\phi=0$. In the following, we are going to show that we can use this remaining gauge freedom to impose stronger gauge conditions.

Notice that $\sqrt{|g|}=\sqrt{G}$. Therefore,
$$\delta A=\pa_{\alpha}A^{\alpha}+\frac{1}{\sqrt{G}}\pa_a\left(\sqrt{G}A^a\right).$$
We want to show that  we can impose    the gauges
\begin{equation}\label{eq2_14_1}
\pa_{\alpha}A^{\alpha}=0,\hspace{1cm}\frac{1}{\sqrt{G}}\pa_a\left(\sqrt{G}A^a\right)=0.
\end{equation}
Any one-form on $M\times \mathcal{N}$ can be written as linear combinations of one-forms of the form:
\begin{equation}\label{eq2_14_2}
A_{\mu}(x,y)dz^{\mu}=q(y)U_{\alpha}(x)dx^{\alpha}+p(x)V_a(y)dy^a,
\end{equation}where $p(x)$ and $q(y)$ are nonzero, but $U_{\alpha}(x)dx^{\alpha}$ or $V_{a}(y)dy^a$ can be zero.
For such a one-form,
\begin{equation}\label{eq1_7_1}
\begin{split}
\pa_{\alpha}A^{\alpha}=&q\delta_M U,\hspace{1cm}\frac{1}{\sqrt{G}}\pa_a\left(\sqrt{G}A^a\right)=p\delta_{\mathcal{N}} V,\\
\delta_{\mathcal{M}} A=&\pa_{\alpha}A^{\alpha}+\frac{1}{\sqrt{G}}\pa_a\left(\sqrt{G}A^a\right)= q\delta_M U +p\delta_{\mathcal{N}} V,
\end{split}\end{equation} where $\delta_{\mathcal{M}}$, $\delta_M$ and $\delta_{\mathcal{N}}$ are the $\delta$-operators on $\mathcal{M}, M$ and $\mathcal{N}$  respectively. In the following, similar conventions will be used for other operators.   Eq. \eqref{eq1_7_1} implies that
  $\delta_{\mathcal{M}}A=0$ if and only if
  $$\frac{\delta_{M}U(x)}{p(x)}=-\frac{  \delta_{\mathcal{N}}V (y)}{q(y)}.$$This happens if and only if there exists a   constant $c$ so that
\begin{align}\label{eq1_6_2}
\delta_{M}U=cp,\quad  \delta_{\mathcal{N}}V =-cq.
\end{align}If $c=0$, then we are done. Otherwise, since
$$F=d_{\mathcal{M}}A=d_{\mathcal{N}}q U +q d_MU+d_Mp V+p d_{\mathcal{N}}V,$$ the equation of motion   gives
\begin{equation}\label{eq1_7_2}
\begin{split}
0=&\delta_{\mathcal{M}} F=\Delta_{\mathcal{N}}q U -\delta_M Ud_{\mathcal{N}}q +q \delta_M d_M U+
\Delta_M p V-d_Mp\delta_{\mathcal{N}}V+p\delta_{\mathcal{N}}d_{\mathcal{N}}V\\
=&\Delta_{\mathcal{N}}q U +q \delta_M d_M U+cq d_Mp+
\Delta_M p V-cp d_{\mathcal{N}}q+p\delta_{\mathcal{N}}d_{\mathcal{N}}V\\
=&\Delta_{\mathcal{N}}q U +q \Delta_M   U+
\Delta_M p V+p\Delta_{\mathcal{N}} V.
\end{split}
\end{equation}Comparing the components, we have
\begin{equation*}
\Delta_{\mathcal{N}}q U +q \Delta_M   U=0,\hspace{1cm}\Delta_M p V+p\Delta_{\mathcal{N}} V=0.
\end{equation*}
Therefore, there must exist  constants $\lambda_1$ and $\lambda_2$ such that
\begin{equation}\label{eq1_7_3}
\frac{\Delta_{\mathcal{N}}q }{q} =-\frac{ \Delta_M   U}{U}=\lambda_1,\hspace{1cm}-\frac{\Delta_M p }{p}=\frac{\Delta_{\mathcal{N}} V}{V}=\lambda_2.
\end{equation} From \eqref{eq1_6_2} and the fact that $\delta_M^2=0$, we find that
\begin{equation*}
\begin{split}
-c\lambda_2p=c\Delta_M p = c\delta_Md_Mp =\delta_Md_M\delta_M U =\delta_M\left(d_M\delta_M+\delta_M d_M\right)U=\delta_M \Delta_M U =-\lambda_1\delta_M U =-c\lambda_1 p.
\end{split}
\end{equation*}This implies that $\lambda_1=\lambda_2$.  Let $\lambda=\lambda_1=\lambda_2$. If $\lambda=0$, then $\Delta_{\mathcal{N}}V=0$ implies that $\delta_{\mathcal{N}}V=0$. Hence $cq=0$. This is a contradiction since we assume that $c\neq 0$ and $q\neq 0$. Therefore $\lambda\neq 0$.
Consider the function $$\phi=\frac{c}{\lambda}pq.$$ It is easy to verify that $\Delta\phi=0$.
Therefore one can use the remaining gauge freedom to transform $A$ to $A'$, where
$$A'=A+d\phi=qU'+pV'=q\left(U+\frac{c}{\lambda}d_Mp\right)+p\left(V+\frac{c}{\lambda}d_{\mathcal{N}}q\right).$$
It follows that
\begin{equation*}\begin{split}
\delta_M U'=&\delta_M\left(U+\frac{c}{\lambda}d_Mp\right)=\delta_MU+\frac{c}{\lambda}\delta_M d_M p=0,\\
\delta_{\mathcal{N}}V'=&\delta_{\mathcal{N}}\left(V+\frac{c}{\lambda}d_{\mathcal{N}}q\right)=\delta_{\mathcal{N}}V+\frac{c}{\lambda}\delta_{\mathcal{N}} d_{\mathcal{N}}q=0,
\end{split}\end{equation*}which show that $A'$ has the desired property \eqref{eq2_14_1}.
   As a conclusion, it is  possible to impose the gauges
\eqref{eq2_14_1} which  are equivalent to
$\delta_{M}U=0, \delta_{\mathcal{N}}V=0$ if $A$ has the form \eqref{eq2_14_2}.   After fixing these gauges, one can show that since $\mathcal{N}$ is assumed to be compact and connected, one only has the gauge freedom of adding to $A$ the differential   of a function $\varphi(x)$ satisfying $\Delta_M\varphi=0$. The gauge condition $\delta_{\mathcal{N}}V=0$ can be considered as a generalization of the almost axial gauge used in \cite{3} when $\mathcal{N}=S^1$.

Before ending this  section, we would like to remark that for general $A=qU+pV$ satisfying $\delta_MU=0$ and $\delta_{\mathcal{N}}V=0$, the equation of motion \eqref{eq1_7_2} still implies \eqref{eq1_7_3}, but in general $\lambda_1\neq \lambda_2$. Therefore, we can separately consider one forms $A$ of the form $qU$, with $$\delta_M U=0, \quad \Delta_M U=-\lambda_1U,\quad \Delta_{\mathcal{N}}q=\lambda_1q,$$ and of the form $pV$ with
$$  \delta_{\mathcal{N}}V=0,\quad \Delta_Mp=-\lambda_2p, \quad\Delta_{\mathcal{N}}V=\lambda_2V.$$
When $\lambda_1=0$, $q$ is a constant and therefore we can further impose the gauge condition $U_0=0$ on $U$.

\section{The eigenmodes of the field inside a cylinder} \label{s3}

As discuss in the previous section, we can consider the eigenmodes of the electromagnetic field of the form
\begin{enumerate}
\item[(I)] $U_{\alpha}dx^{\alpha}$ with $U_0=0$ and $\delta_MU=0$;
\item[(II)] $q_j(y)U_{\alpha}(x)dx^{\alpha}$ with $\delta_MU=0$, $\Delta_MU+m_j^2U=0$, $j=1,2,\ldots$, where $q_j(y)$ is an eigenfunction with nonzero eigenvalue $m_j^2$   of the Laplace operator on functions on $\mathcal{N}$;
\item[(III)] $p(x)V_{j,a}(y)dy^a$ with $\Delta_M p+\mu_j^2 p =0$, $j=1,2,\ldots$, where $V_{j,a}(y)dy^a$ is a co-closed eigen-one-form with eigenvalue $\mu_j^2$   of the Laplace operator  on  $\mathcal{N}$.\end{enumerate}

In the following, we  find the eigenmodes of the electromagnetic field in the cylinder $[0,a]\times \Omega\times\mathcal{N}$ with combinations of perfectly conducting or infinitely permeable boundary conditions on the sidewall $[0,a]\times\pa\Omega\times \mathcal{N}$, the bottom $x^1=0$ and the top $x^1=a$.

 The eigenmodes   can   be divided into TE modes which are modes with $F_{01}=0$, and TM modes which are modes with $F_{\mu\nu}=0$ for all $\mu,\nu\neq 0, 1$.
Denote by $\bar{x}=(x^2, x^{3})$. Let $ \varphi_1(\bar{x}),\varphi_2(\bar{x}),\ldots$ be the eigenfunctions of the Laplace operator with Dirichlet boundary conditions on $\Omega$, with eigenvalues $\varpi_{1}^2,\varpi_{2}^2,\ldots$; and let $ \psi_0(\bar{x}), \psi_1(\bar{x}),\psi_2(\bar{x}),\ldots$ be the   eigenfunctions of the Laplace operator with Neumann boundary conditions on $\Omega$, with eigenvalues $  \varkappa_{0,}^2, \varkappa_{1}^2,\varkappa_2^2,\ldots$. $\psi_0(\bar{x})$ is the constant function with eigenvalue $\varkappa_0^2=0$.

\subsection{Perfectly conducting condition on the whole cylinder }\label{s3_1}

When the whole cylinder is perfectly conducting \eqref{eq2_21_1}, one can show that the set of eigenmodes of the electromagnetic field is given by

\vspace{0.3cm}\noindent\textbf{Type A TE modes:}
\begin{equation*}
\begin{split}
\begin{aligned}
&A_2=-\sin\frac{\pi k x}{a}\frac{\pa \psi_{l}(\bar{x})}{\pa x^{3}}e^{-i\omega t}q_j(y),\\
&A_{3}=\sin\frac{\pi k x}{a}\frac{\pa \psi_{l}(\bar{x})}{\pa x^2}e^{-i\omega t}q_j(y),\\
&\text{all other $A_{\mu}=0$},\end{aligned}\hspace{2cm} \begin{aligned}&\omega^2=\left(\frac{\pi k}{a}\right)^2+\varkappa_l^2+m_j^2,\\&k,l=1,2,\ldots;\;j=0,1,2,\ldots.\end{aligned}
\end{split}
\end{equation*}The $j=0$ modes are type I modes, and the $j\geq 1$ modes are type II modes. By convention, $m_0=0$ and $q_0(y)=1$.

\vspace{0.3cm}\noindent\textbf{Type B TE modes:} These include all the type III modes where
\begin{equation*}
\begin{split}
\begin{aligned}
&A_{\alpha}=0,\\
&A_a=\sin\frac{\pi k x}{a}\varphi_{l}(\bar{x})e^{-i\omega t}V_{j,a}(y),\end{aligned}\hspace{2cm} \begin{aligned} \omega^2=&\left(\frac{\pi k}{a}\right)^2+\varpi_l^2+\mu_j^2,\\ &k,l,j=1,2,\ldots.\end{aligned}
\end{split}
\end{equation*}

\vspace{0.3cm}\noindent\textbf{Type A TM modes:}
\begin{equation*}
\begin{split}
\begin{aligned}&A_0=A_a=0, \\
&A_1=\varpi_l^2\cos\frac{\pi k x}{a}\varphi_{l}(\bar{x})e^{-i\omega t}q_j(y),\\
& A_{\gamma}=-\frac{\pi k}{a}\sin\frac{\pi k x}{a}\frac{\pa \varphi_{l}(\bar{x})}{\pa x^{\gamma}}e^{-i\omega t}q_j(y),\quad  \gamma=2,3, \end{aligned}\hspace{2cm}\begin{aligned}&\omega^2=\left(\frac{\pi k}{a}\right)^2+\varpi_l^2+m_j^2, \\&k,j=0,1,2,\ldots;\;l=1,2,\ldots.\end{aligned}
\end{split}\end{equation*}The $j=0$ modes are type I modes, and the $j\geq 1$ modes are type II modes.

\vspace{0.3cm} \noindent\textbf{Type B TM modes:}  These are   type II modes with
\begin{equation*}
\begin{split}
\begin{aligned}
&A_1=A_a=0,\\
&A_0=\varpi_l^2\sin\frac{\pi k x}{a}\varphi_{l}(\bar{x})e^{-i\omega t}q_j(y),\\
&A_{\gamma}=i\omega\sin\frac{\pi k x}{a}\frac{\pa \varphi_{l}(\bar{x})}{\pa x^{\gamma}}e^{-i\omega t}q_j(y),\quad   \gamma=2,3,
\end{aligned}\hspace{2cm} \begin{aligned}\omega^2=&\left(\frac{\pi k}{a}\right)^2+\varpi_l^2+m_j^2,\\
&k,l,j=1,2,\ldots.\end{aligned}
\end{split}
\end{equation*}

\subsection{Infinitely permeable condition on the whole cylinder}\label{s3_2}

When the whole cylinder is infinitely permeable \eqref{eq2_21_2}, one can show that the set of eigenmodes of the electromagnetic field is given by

\vspace{0.3cm}\noindent\textbf{ Type A TE modes:}
\begin{equation*}
\begin{split}
\begin{aligned}
&A_2=-\cos\frac{\pi k x}{a}\frac{\pa \varphi_{l}(\bar{x})}{\pa x^{3}}e^{-i\omega t}q_j(y),\\
&A_{3}=\cos\frac{\pi k x}{a}\frac{\pa \varphi_{l}(\bar{x})}{\pa x^2}e^{-i\omega t}q_j(y),\\
&\text{all other $A_{\mu}=0$},\end{aligned}\hspace{2cm} \begin{aligned}&\omega^2=\left(\frac{\pi k}{a}\right)^2+\varpi_l^2+m_j^2,\\&l=1,2,\ldots;\;k,j=0,1,2,\ldots.\end{aligned}
\end{split}
\end{equation*}

\vspace{0.3cm}\noindent\textbf{ Type B TE modes:}
\begin{equation*}
\begin{split}
\begin{aligned}
&A_{\alpha}=0,\\
&A_a=\cos\frac{\pi k x}{a}\psi_{l}(\bar{x})e^{-i\omega t}V_{j,a}(y),\end{aligned}\hspace{2cm} \begin{aligned} &\omega^2=\left(\frac{\pi k}{a}\right)^2+\varkappa_l^2+\mu_j^2,\\ &k,l=0,1,2,\ldots;\;j=1,2,\ldots.\end{aligned}
\end{split}
\end{equation*}Notice that  the space of one-forms $V$ on  $\mathcal{N}$  with $\delta_{\mathcal{N}}V=0$ contains harmonic one-forms where $\Delta_{\mathcal{N}}V=0$. Let $h$ denotes the first Betti number of $\mathcal{N}$ -- the dimension of the vector space of harmonic one-forms on $\mathcal{N}$, which is a topological invariant. Then the set of $\mu_j^2$ contains $h$ zeros. Without loss of generality, let $\mu_1^2,\ldots,\mu_h^2$ be equal to zero. Then the modes with $l=0$ and $j=1,2,\ldots,h$ are also TM modes. Therefore, they are TEM modes.

\vspace{0.3cm}\noindent\textbf{ Type A TM modes:}
\begin{equation*}
\begin{split}
\begin{aligned}&A_0=A_a=0, \\
&A_1=\varkappa_l^2\sin\frac{\pi k x}{a}\psi_{l}(\bar{x})e^{-i\omega t}q_j(y),\\
& A_{\gamma}=\frac{\pi k}{a}\cos\frac{\pi k x}{a}\frac{\pa \psi_{l}(\bar{x})}{\pa x^{\gamma}}e^{-i\omega t}q_j(y),\quad  \gamma=2,3,
\end{aligned}\hspace{2cm}\begin{aligned}&\omega^2=\left(\frac{\pi k}{a}\right)^2+\varkappa_l^2+m_j^2, \\&j=0,1,2,\ldots;\;k,l=1,2,\ldots.  \end{aligned}
\end{split}\end{equation*}

\vspace{0.3cm} \noindent\textbf{ Type B TM modes:}  These are   type II modes with
\begin{equation*}
\begin{split}
\begin{aligned}
&A_1=A_a=0,\\
&A_0=\varkappa_l^2\cos\frac{\pi k x}{a}\psi_{l}(\bar{x})e^{-i\omega t}q_j(y)\\
&A_{\gamma}=i\omega\cos\frac{\pi k x}{a}\frac{\pa \psi_{l}(\bar{x})}{\pa x^{\gamma}}e^{-i\omega t}q_j(y),\quad   \gamma=2,3,\end{aligned}\hspace{2cm} \begin{aligned}&\omega^2=\left(\frac{\pi k}{a}\right)^2+\varkappa_l^2+m_j^2,\\
&k=0,1,2,\ldots;\;l,j=1,2,\ldots.\end{aligned}
\end{split}
\end{equation*}or
\begin{equation*}
\begin{split}
\begin{aligned}
&A_0=\frac{\pi k}{a}\cos\frac{\pi k x}{a} e^{-i\omega t}q_j(y),\\
&A_1=-i\omega\sin\frac{\pi k x}{a} e^{-i\omega t}q_j(y),\\
&\text{all other $A_{\mu}=0$},\end{aligned}\hspace{4cm} \begin{aligned} \omega^2=&\left(\frac{\pi k}{a}\right)^2+m_j^2,\\
&k,j= 1,2,\ldots.\end{aligned}
\end{split}
\end{equation*}

\subsection{Perfectly conducting condition on the sidewall and the bottom,   infinitely permeable condition on the top }\label{s3_3}

When the side wall and the bottom of the cylinder are perfectly conducting \eqref{eq2_21_1}, and the top is infinitely permeable \eqref{eq2_21_2}, it is immediate to check that the eigenmodes of the electromagnetic field are obtained by replacing the $k$ for the modes in Section \ref{s3_1}   by $\displaystyle k+\frac{1}{2}$, where $k$ runs from zero to infinity.

\subsection{Infinitely permeable condition on the sidewall and the bottom,   perfectly conducting condition on the top}\label{s3_4}

When the side wall and the bottom of the cylinder are infinitely permeable \eqref{eq2_21_2}, and the top is perfectly conducting \eqref{eq2_21_1}, it is immediate to check that the eigenmodes of the electromagnetic field  are obtained by replacing the $k$ for the modes in Section \ref{s3_2}   by $\displaystyle k+\frac{1}{2}$, where $k$ runs from zero to infinity.

\vspace{0.5cm} Note that in the absence of the space $\mathcal{N}$, we only have the type A TE modes with $j=0$ and the type A TM modes with $j=0$. The   type B TE modes and type B TM modes only exist in the presence of the space $\mathcal{N}$.

\section{The Casimir energy and the Casimir force}\label{s4}

\subsection{The cylinder and the piston are imposed with the same boundary conditions }\label{s4_1} When the cylinder and the piston are both perfectly conducting, the piston divides the cylinder $[0,L]\times\Omega\times\mathcal{N}$ into two cylinders $[0,a]\times \Omega\times\mathcal{N}$ and $[0,L-a]\times \Omega\times\mathcal{N}$, both of them are perfectly conducting everywhere. From the results of Section \ref{s3_1}, we find that  the zeta function
$\zeta_T(s;a)$ is given by
\begin{equation*}\begin{split}
\zeta_T(s;a)=& \sum_{p=-\infty}^{\infty}\sum_{k=1}^{\infty}\sum_{l=1}^{\infty}\sum_{j=0}^{\infty}\left(\left[\frac{\pi k}{a}\right]^2+\varkappa_l^2+m_j^2+[2\pi p T]^2\right)^{-s}
+\sum_{p=-\infty}^{\infty}\sum_{k=1}^{\infty}\sum_{l=1}^{\infty}\sum_{j=1}^{\infty}\left(\left[\frac{\pi k}{a}\right]^2+\varpi_l^2+\mu_j^2+[2\pi p T]^2\right)^{-s}\\
&+\sum_{p=-\infty}^{\infty}\sum_{k=0}^{\infty}\sum_{l=1}^{\infty}\sum_{j=0}^{\infty}\left(\left[\frac{\pi k}{a}\right]^2+\varpi_l^2+m_j^2+[2\pi p T]^2\right)^{-s}+
\sum_{p=-\infty}^{\infty}\sum_{k=1}^{\infty}\sum_{l=1}^{\infty}\sum_{j=1}^{\infty}\left(\left[\frac{\pi k}{a}\right]^2+\varpi_l^2+m_j^2+[2\pi p T]^2\right)^{-s}.
\end{split}\end{equation*}  For the third term, the $k=0$ terms    do not depend on $a$. Therefore, the zeta function can be written as
\begin{equation}\label{eq1_25_4}\begin{split}
\zeta_T(s;a)=&\sum_{p=-\infty}^{\infty} \sum_{\alpha} \zeta_{\alpha,p}(s)+\mathcal{C}(s),\\
\zeta_{\alpha,p}(s;a):=&\sum_{k=1}^{\infty}  \left(\left[\frac{\pi k}{a}\right]^2+\tau_{\alpha,p}^2\right)^{-s},
\end{split}\end{equation}where $\mathcal{C}(s)$ denotes a term independent of $a$ whose value can change from one expression to another, and $$\tau_{\alpha,p}^2=\tau_{\alpha}^2+(2\pi pT)^2.$$The set of $\tau_{\alpha}^2$ contains:

\begin{enumerate}\item[$-$]
$\varpi_l^2+m_j^2$, $j\geq 0$, $l\geq 1$, with multiplicity two if $j\neq 0$ and multiplicity one if $j=0$,
\item[$-$] $\varpi_l^2+\mu_j^2$, $j\geq 1, l\geq1$, each with multiplicity one, \hfill (PC)
\item[$-$] $\varkappa_l^2+m_j^2$, $j\geq 0, l\geq 1$, each with multiplicity one.
\end{enumerate}Notice that none of these $\tau_{\alpha}^2$ is zero.

When the cylinder and the piston are both infinitely permeable, we have two infinitely permeable cylinders.  The results in Section \ref{s3_2} show that  the zeta function $\zeta_T(s;a)$ can also be written in the form \eqref{eq1_25_4}, where
the set of $\tau_{\alpha}^2$ contains:

\begin{enumerate}\item[$-$]
$\varkappa_l^2+m_j^2$, $j\geq 0$, $l\geq 1$, with multiplicity two if $j\neq 0$ and multiplicity one if $j=0$,
\item[$-$] $\varkappa_l^2+\mu_j^2$, $j\geq 1, l\geq 0$, each with multiplicity one, \hfill(IP)
\item[$-$] $\varpi_l^2+m_j^2$, $j\geq 0, l\geq 1$, each with multiplicity one,
\item[$-$] $m_j^2$, $j\geq  1$, each with multiplicity one.
\end{enumerate}In this case, we find that there are $h$ of the $\tau_{\alpha}^2$ that are equal to zero, which are the $\tau_{\alpha}^2$ corresponding to the TEM modes, i.e., $\varkappa_0^2+\mu_1^2,\ldots,\varkappa_0^2+\mu_h^2$.

Using the fact that
\begin{equation*}
\sum_{k=1}^{\infty}\exp\left(-t\left[\frac{\pi k}{a}\right]^2\right)=-\frac{1}{2}+\frac{a}{2\sqrt{\pi}}t^{-\frac{1}{2}}+\frac{a}{\sqrt{\pi}}t^{-\frac{1}{2}}\sum_{k=1}^{\infty} \exp\left(-\frac{k^2a^2}{t}\right),
\end{equation*}we find that if $\tau_{\alpha,p}^2\neq 0$,
\begin{equation*}
\begin{split}
\zeta_{\alpha,p}(s;a)=& \frac{1}{\Gamma(s)} \sum_{k=1}^{\infty}  \int_0^{\infty}t^{s-1}\exp\left\{-t\left(\left[\frac{\pi k}{a}\right]^2+\tau_{\alpha,p}^2\right)\right\}dt\\
=&\mathcal{C}_{\alpha,p}(s)+a\mathcal{D}_{\alpha,p}(s)+\frac{2a}{\sqrt{\pi}\Gamma(s)} \sum_{k=1}^{\infty} \left(\frac{ka}{\tau_{\alpha,p}}\right)^{s-\frac{1}{2}}K_{s-\frac{1}{2}}\left(2ka\tau_{\alpha,p}\right).
\end{split}
\end{equation*}Here $\mathcal{C}_{\alpha,p}(s)$ and $\mathcal{D}_{\alpha,p}(s)$ are  terms independent of $a$.
From this, we obtain
\begin{equation*}
\begin{split}
\zeta_{\alpha,p}(0;a)=&\mathcal{C}_{\alpha,p}(0)+a\mathcal{D}_{\alpha,p}(0),\\
\zeta_{\alpha,p}'(0;a)=&\mathcal{C}_{\alpha,p}'(0)+a\mathcal{D}_{\alpha,p}'(0)+\frac{2a}{\sqrt{\pi} } \sum_{k=1}^{\infty} \left(\frac{\tau_{\alpha,p}}{ka}\right)^{ \frac{1}{2}}K_{ \frac{1}{2}}\left(2ka\tau_{\alpha,p}\right)\\
=&\mathcal{C}_{\alpha,p}'(0)+a\mathcal{D}_{\alpha,p}'(0)+  \sum_{k=1}^{\infty} \frac{1}{k}e^{-2ka\tau_{\alpha,p}}.
\end{split}\end{equation*}
On the other hand, if $\tau_{\alpha,p}^2=0$, $$\zeta_{\alpha,p}(s)=\left(\frac{\pi}{a}\right)^{-2s}\zeta_R(2s).$$ It follows that
\begin{equation*}
\begin{split}
\zeta_{\alpha,p}(0;a)=&\zeta_R(0)=-\frac{1}{2}\\
\zeta_{\alpha,p}'(0;a)=&2\zeta_R'(0)-2\zeta_R(0)\log \frac{\pi}{a}=-\log(2\pi)+\log\frac{\pi}{a}=-\log(2a)
\end{split}\end{equation*}Here $\zeta_R(s)$ is the Riemann zeta function.

From \eqref{eq1_24_2}, we find that  the Casimir energy of the piston system \eqref{eq1_24_1} is given by
\begin{equation*}\begin{split}
E_{\text{Cas}}^{\text{piston}}=&\mathcal{E}_0-\frac{T}{2} \sum_{k=1}^{\infty}\sum_{\tau_{\alpha,p}\neq 0}\frac{1}{k}e^{-2ka\tau_{\alpha,p}}
+\frac{T}{2}\sum_{\tau_{\alpha,p}=0}\log(\lambda a)\\&-\frac{T}{2} \sum_{k=1}^{\infty}\sum_{\tau_{\alpha,p}\neq 0}\frac{1}{k}e^{-2k(L-a)\tau_{\alpha,p}} +\frac{T}{2}\sum_{\tau_{\alpha,p}=0}\log(\lambda (L-a)),
\end{split}
\end{equation*}where $\mathcal{E}_0$ is independent of $a$. It follows that the Casimir force acting on the piston is given by
\begin{equation}\label{eq1_25_2}
F_{\text{Cas}}^{\text{piston}}=-\frac{\pa E_{\text{Cas}}^{\text{piston}}}{\pa a}=F_{\text{Cas}}^{\parallel}(a)-F_{\text{Cas}}^{\parallel}(L-a),
\end{equation}where
\begin{equation}\label{eq1_25_1}\begin{split}
F_{\text{Cas}}^{\parallel}(a)=& -\frac{\pa  }{\pa a}\left(-\frac{T}{2} \sum_{k=1}^{\infty}\sum_{\tau_{\alpha,p}\neq 0}\frac{1}{k}e^{-2ka\tau_{\alpha,p}}
+\frac{T}{2}\sum_{\tau_{\alpha,p}=0}\log(\lambda a)\right)
\\=&-T \sum_{k=1}^{\infty}\sum_{\tau_{\alpha,p}\neq 0}\tau_{\alpha,p}e^{-2ka\tau_{\alpha,p}}-\sum_{\tau_{\alpha,p}=0}\frac{T}{2a}=
-T \sum_{\tau_{\alpha,p}\neq 0}\frac{\tau_{\alpha,p}}{e^{2a\tau_{\alpha,p}}-1}-\sum_{\tau_{\alpha,p}=0}\frac{T}{2a}.
\end{split}\end{equation}In fact, $F_{\text{Cas}}^{\parallel}(a)$ is   the limit of the Casimir force acting on the piston $F_{\text{Cas}}^{\text{piston}}$ when the cylinder becomes infinitely long, i.e., $L\rightarrow\infty$. Therefore, it is the Casimir force acting between two perfectly conducting plates or two infinitely permeable plates separated by a distance $a$ moving inside an infinitely long cylinder with cross section $\Omega\times \mathcal{N}$.

When both the cylinder and the piston are infinitely permeable, the last term in \eqref{eq1_25_1} which comes from the TEM modes is nonzero if $h$, the dimension of the space of harmonic one-forms of $\mathcal{N}$ is nonzero. In contrast to the contribution to the force from the terms with $\tau_{\alpha,p}\neq 0$ which decays to zero exponentially fast when $a\rightarrow\infty$, the contribution from the terms with $\tau_{\alpha,p}=0$ has power law decay which is much slower. This gives rise to a long range Casimir force \cite{9}. It is interesting to note that the long range term $\displaystyle -\frac{T}{2a}$ is a limit of $\displaystyle
-T\frac{\tau_{\alpha,p}}{e^{2a\tau_{\alpha,p}}-1}$ when $\tau_{\alpha,p}\rightarrow 0$, which exhibits some kind of smooth transition.

From the expression \eqref{eq1_25_1}, it is obvious that the Casimir force acting between two perfectly conducting plates or two infinitely permeable plates is always attractive, and is a monotonically decreasing function of the distance between the plates. Therefore, it can be inferred from \eqref{eq1_25_2} that in a closed   cylinder of finite length, the Casimir force acting on the piston which has the same boundary condition with the cylinder always tends to pull the   piston to the closer end.

In the following, we study the asymptotic behavior of the Casimir force \eqref{eq1_25_1} at different limits. Denote by $r=\sqrt[n]{ \mathcal{V}(\mathcal{N})}$ a measure of the size of the manifold $\mathcal{N}$ and $R=\sqrt{\mathcal{A}(\Omega)}$ a measure of the size of the domain $\Omega$. We will investigate the behavior of the Casimir force when the length scales $a, r$ and $R$ are such that $r\ll a\ll R$ or $a\ll r, R$, and when $aT\ll 1$ or $aT\gg 1$.

 \eqref{eq1_25_1} is the high temperature expansion of the Casimir force. It shows that when $aT\gg 1$, the Casimir force is dominated by the term
$$F_{\text{Cas}}^{\parallel}(a)\sim -T\sum_{\tau_\alpha\neq 0}\frac{\tau_{\alpha}}{e^{2a\tau_{\alpha}}-1}-\sum_{ \tau_{\alpha} = 0}\frac{T}{2a},$$which is linear in $T$. This term is called the classical term. The sum of the remaining terms decay exponentially.

In Appendix \ref{a_1}, we show that the Casimir force \eqref{eq1_25_1} can be rewritten as
\begin{equation*}
\begin{split}
F_{\text{Cas}}^{\parallel}(a)
=&-\sum_{\tau_{\alpha}=0}\left\{\frac{\pi}{24a^2}+\frac{\pi T^2}{6} \right\}-\sum_{\tau_{\alpha}\neq 0}\left\{\frac{1}{2\pi a}\sum_{k=1}^{\infty}\frac{\tau_{\alpha}}{k} K_1(2ka\tau_{\alpha})+\frac{1}{\pi}\sum_{k=1}^{\infty} \tau_{\alpha}^2K_0(2ka\tau_{\alpha})
+\frac{T}{\pi}\sum_{p=1}^{\infty}\frac{\tau_{\alpha}}{p}K_1\left(\frac{p\tau_{\alpha}}{T}\right)\right\}\\&+\frac{\pi^2}{a^3}
\sum_{\tau_{\alpha}}\sum_{k=1}^{\infty}
\frac{k^2}{\sqrt{\left[\frac{\pi k}{a}\right]^2+\tau_{\alpha}^2}}\left(\exp\left(\frac{1}{T}\sqrt{\left[\frac{\pi k}{a}\right]^2+\tau_{\alpha}^2} \right) -1\right)^{-1}.
\end{split}
\end{equation*}This shows that at zero temperature, the Casimir force is given by
\begin{equation*}
\begin{split}
F_{\text{Cas}}^{\parallel, T=0}(a)
=&-\sum_{\tau_{\alpha}=0}  \frac{\pi}{24a^2} -\sum_{\tau_{\alpha}\neq 0}\left\{\frac{1}{2\pi a}\sum_{k=1}^{\infty}\frac{\tau_{\alpha}}{k} K_1(2ka\tau_{\alpha})+\frac{1}{\pi}\sum_{k=1}^{\infty} \tau_{\alpha}^2K_0(2ka\tau_{\alpha})
 \right\}.\end{split}
\end{equation*}In the case the two   infinitely permeable plates are placed inside an infinitely permeable cylinder and the first Betti number $h$ of $\mathcal{N}$ is nonzero, the leading term of the thermal correction is
$$-\frac{\pi h}{6}T^2.$$ Otherwise, the thermal correction goes to zero exponentially fast when $aT\rightarrow 0$.

 Next we consider the behavior of the Casimir force \eqref{eq1_25_1} when  the separation between the plates $a$ is much smaller than the size $R$ of the domain $\Omega$, i.e., $a\ll R$. In this case, the result of Appendix \ref{b} shows that  the first two leading terms of the Casimir force is given by
\begin{equation}\label{eq2_11_1}
\begin{split}
F_{\text{Cas}}^{\parallel} (a)\sim &  \mathcal{A}(\Omega)\left\{-(2+h)\frac{T }{8\pi a^3}\zeta_R(3)-\frac{T }{8\pi}
\sum_{k=1}^{\infty}\sum_{\sigma_{\beta,p}\neq 0} \left(\frac{2\sigma_{\beta,p}^2}{ka}+\frac{2\sigma_{\beta,p}}{k^2a^2}+\frac{1}{k^3a^3}\right)e^{-2ka\sigma_{\beta,p}}\right\}
\\&\pm l(\pa\Omega) \left\{h\frac{\pi T}{96 a^2} +\frac{T}{8\pi}\sum_{k=1}^{\infty}\sum_{\xi_{\gamma,p}\neq 0} \frac{\xi_{\gamma,p}}{ka}K_1(2ka\xi_{\gamma,p})+\frac{T}{4\pi}\sum_{k=1}^{\infty}\sum_{\xi_{\gamma,p}\neq 0} \xi_{\gamma,p}^2K_0(2ka\xi_{\gamma,p})\right\},
 \end{split}
\end{equation}or
\begin{equation}\label{eq2_11_2}
\begin{split}
F_{\text{Cas}}^{\parallel} (a)\sim &  \mathcal{A}(\Omega)\left\{-(2+h) \left(\frac{\pi^2}{480a^4} +\frac{  \pi^2T^4 }{90}\right)
-\frac{3 }{8\pi^2}\sum_{k=1}^{\infty}\sum_{\sigma_{\beta}\neq 0}e^{2\pi i k\chi}\left(\frac{\sigma_{\beta}}{ka}\right)^2K_2(2ka\sigma_{\beta})
-\frac{ 1}{4\pi^2}\sum_{k=1}^{\infty}\sum_{\sigma_{\beta}\neq 0}   \frac{\sigma_{\beta}^3}{ka} K_1(2ka\sigma_{\beta})
\right.\\&\left. -\frac{ T^2}{2\pi^2}\sum_{p=1}^{\infty}
\sum_{\sigma_{\beta}\neq 0}\left(\frac{\sigma_{\beta}}{p}\right)^2 K_2\left(\frac{p\sigma_{\beta}}{T}\right)
 +\frac{\pi  T}{2a^3}\sum_{p=1}^{\infty}\sum_{k=1}^{\infty}\sum_{\sigma_{\beta}}\frac{k^2}{p}
\exp\left(-\frac{p}{T}\sqrt{\left[\frac{\pi k}{a}\right]^2+\sigma_{\beta}^2}\right)\right\}
\\&\pm l(\pa\Omega) \left\{h \left(\frac{\zeta_R(3)}{32\pi a^3} +\frac{\zeta_R(3)T^3}{8\pi}\right)+\frac{1}{32\pi}
\sum_{k=1}^{\infty}\sum_{\xi_{\gamma}\neq 0} e^{-2ka\xi_{\gamma}}\left(\frac{2\xi_{\gamma}^2}{ka}+\frac{2\xi_{\gamma}}{k^2a^2}+\frac{1}{k^3a^3}\right)
\right.\\&\left.+\frac{ T^2}{8\pi}\sum_{p=1}^{\infty}\sum_{\xi_{\gamma}\neq 0}e^{-\frac{p\xi_{\gamma}}{T}}\left(\frac{\xi_{\gamma}}{p^2}+\frac{T}{p^3}\right)-\frac{ \pi}{4a^3}\sum_{k=1}^{\infty}\sum_{p=1}^{\infty}\sum_{\xi_{\gamma}}k^2K_0\left(\frac{p}{T}\sqrt{\left[\frac{\pi k}{a}\right]^2+\xi_{\gamma}^2}\right)\right\},
 \end{split}
\end{equation}
where $\sigma_{\beta,p}^2=\sigma_{\beta}^2+[2\pi p T]^2$, $\xi_{\gamma,p}^2=\xi_{\gamma}^2+[2\pi p T]^2$; $A(\Omega)$ is the area of $\Omega$ which is of order $R^2$, and $l(\pa\Omega)$ is the length of the boundary $\pa\Omega$ of $\Omega$ which is of order $R$. The set of $\sigma_{\beta}^2$ contains $m_0^2$ with multiplicity two, $m_j^2, j\geq 1$, with multiplicity three, and $\mu_j^2, j\geq 1,$ with multiplicity one; the set of $\xi_{\gamma}^2$ contains   $m_j^2, j\geq 1$,   and $\mu_j^2, j\geq 1$, each with multiplicity one. The plus signs on the second line in \eqref{eq2_11_1}  and the third line in \eqref{eq2_11_2} are for the case where the cylinder is perfectly conducting, and the minus signs are for the case where the cylinder is infinitely permeable.

In the high temperature regime, i.e., $aT\gg 1$,  the leading terms of the Casimir force when $a\ll R$ is given by the sum of the $p=0$ terms in \eqref{eq2_11_1}. In the low temperature regime, i.e., $aT\ll 1$, \eqref{eq2_11_2} shows that the leading terms of the Casimir force is given by
\begin{equation}\label{eq2_11_2_2}
\begin{split}
F_{\text{Cas}}^{\parallel} (a)\sim &  \mathcal{A}(\Omega)\left\{-(2+h)  \frac{\pi^2}{480a^4}
-\frac{3 }{8\pi^2}\sum_{k=1}^{\infty}\sum_{\sigma_{\beta}\neq 0}e^{2\pi i k\chi}\left(\frac{\sigma_{\beta}}{ka}\right)^2K_2(2ka\sigma_{\beta})
-\frac{ 1}{4\pi^2}\sum_{k=1}^{\infty}\sum_{\sigma_{\beta}\neq 0}   \frac{\sigma_{\beta}^3}{ka} K_1(2ka\sigma_{\beta})
 \right\}
\\&\pm l(\pa\Omega) \left\{h \frac{\zeta_R(3)}{32\pi a^3}  +\frac{1}{32\pi}
\sum_{k=1}^{\infty}\sum_{\xi_{\gamma}\neq 0} e^{-2ka\xi_{\gamma}}\left(\frac{2\xi_{\gamma}^2}{ka}+\frac{2\xi_{\gamma}}{k^2a^2}+\frac{1}{k^3a^3}\right)
 \right\}\\
 &-(2+h)\mathcal{A}(\Omega)\frac{\pi^2T^4}{90}\pm h l(\pa\Omega) \frac{\zeta_R(3)T^3}{8\pi}.
 \end{split}
\end{equation}The first two lines give the zero temperature contribution, and the last line gives the thermal correction which are of polynomial order in $T$. The remaining terms go to zero exponentially fast when $aT\rightarrow 0$.

\eqref{eq2_11_1} and \eqref{eq2_11_2} can also be used to study the leading behavior of the Casimir force when  $r\ll a \ll R$. Since $\sigma_{\beta}$ and $\xi_{\gamma}$ are proportional to $r^{-1}$, we find that in the limit $r/a\rightarrow 0$, the leading term of the Casimir force is given by
\begin{equation}\label{eq2_11_5}
\begin{split}
F_{\text{Cas}}^{\parallel} (a)\sim & \frac{2+h}{2} \mathcal{A}(\Omega)\left\{- \frac{T }{4\pi a^3}\zeta_R(3)-\frac{T }{\pi}
\sum_{k=1}^{\infty}\sum_{p=1}^{\infty} \left(\frac{4\pi^2p^2T^2}{ka}+\frac{2\pi p T}{k^2a^2}+\frac{1}{2k^3a^3}\right)e^{-4\pi kpaT}\right\},
\end{split}
\end{equation}or
\begin{equation}\label{eq2_11_6}
\begin{split}
F_{\text{Cas}}^{\parallel} (a)\sim &  \frac{2+h}{2}\mathcal{A}(\Omega)\left\{-  \frac{\pi^2}{240a^4} -\frac{  \pi^2T^4 }{45}
 +\frac{\pi  T}{a^3}\sum_{p=1}^{\infty}\sum_{k=1}^{\infty} \frac{k^2}{p}
\exp\left(-\frac{\pi kp}{aT} \right)\right\}.
 \end{split}
\end{equation}
Notice that the expressions in the brackets of \eqref{eq2_11_5} and \eqref{eq2_11_6} are the Casimir force per unit area acting on a pair of perfectly conducting or infinitely permeable plates in the $4D$ Minkowski spacetime \cite{13}. They are also equal to twice the Casimir force per unit area acting on a pair of Dirichlet or Neumann plates \cite{13}. Therefore in the limit the size of the manifold $\mathcal{N}$ goes to zero, one recovers the Casimir force between a pair of infinite parallel plates in the $4D$ Minkowski spacetime if and only if $h=0$, i.e., the first Betti number of $\mathcal{N}$ is zero. For general $h$, one finds that when the size of $\mathcal{N}$ goes to zero, one has $h$ extra copies of Casimir force acting on a pair of Dirichlet plates.  Recall that $h$ is the number of  zero modes for the Laplace operator on one-forms on $\mathcal{N}$. Therefore, the presence of extra $h$ copies of the Casimir force on a pair of Dirichlet  plates when the size of the manifold $\mathcal{N}$ goes to zero can be considered as a kind of instantonic effect.

When $aT\gg 1$, \eqref{eq2_11_5} shows that the leading term of the Casimir force is
\begin{equation}\label{eq2_11_3}
\begin{split}
F_{\text{Cas}}^{\parallel} (a)\sim & -\frac{2+h}{2} \mathcal{A}(\Omega) \frac{T }{4\pi a^3}\zeta_R(3).
 \end{split}
\end{equation}When $aT\ll 1$, \eqref{eq2_11_6} shows that the leading term of the Casimir force is
\begin{equation}\label{eq2_11_4}
\begin{split}
F_{\text{Cas}}^{\parallel} (a)\sim &  -\frac{2+h}{2}\mathcal{A}(\Omega)   \frac{\pi^2}{240a^4}.
 \end{split}
\end{equation}  Up to the factor $(1+h/2)$, these are the familiar leading behavior of the Casimir force acting on a pair of perfectly conducting plates in the $4D$ Minkowski spacetime in the high temperature regime and in the low temperature regime respectively.

Finally, we consider the case where separation between the plates is much smaller than the sizes of both $\Omega$ and $\mathcal{N}$, i.e., $a\ll r, R$. In this case, the results of Appendix \ref{b} shows that the first two leading terms of the Casimir force is given by
\begin{equation}\label{eq2_11_7}
\begin{split}
F_{\text{Cas}}^{\parallel} (a)\sim (n+2) \mathcal{V}(\mathcal{S})&\left\{-T\frac{(n+2)\Gamma\left(\frac{n+3}{2}\right)}{(4\pi)^{\frac{n+3}{2}}a^{n+3}}\zeta_R(n+3)
-(n+2)\frac{T^{\frac{n+5}{2}}}{2^{\frac{n-1}{2}}}\sum_{k=1}^{\infty}\sum_{p=1}^{\infty} \left(\frac{p}{ka}\right)^{\frac{n+3}{2}}K_{\frac{n+3}{2}}(4\pi kp aT)\right.\\
&\left.-\frac{\pi T^{\frac{n+7}{2}}}{2^{\frac{n-5}{2}}}\sum_{k=1}^{\infty}\sum_{p=1}^{\infty} \frac{p^{\frac{n+5}{2}}}{(ka)^{\frac{n+1}{2}}}K_{\frac{n+1}{2}}(4\pi kp aT)\right\}
\\
 \mp\frac{n\mathcal{V}(\pa\mathcal{S})}{4}&\left\{-T\frac{(n+1)\Gamma\left(\frac{n+2}{2}\right)}{(4\pi)^{\frac{n+2}{2}}a^{n+2}}\zeta_R(n+2)
 -(n+1)\frac{T^{\frac{n+4}{2}}}{2^{\frac{n-2}{2}}}\sum_{k=1}^{\infty}\sum_{p=1}^{\infty} \left(\frac{p}{ka}\right)^{\frac{n+2}{2}}K_{\frac{n+2}{2}}(4\pi kp aT)\right.\\
&\left.-\frac{\pi T^{\frac{n+6}{2}}}{2^{\frac{n-6}{2}}}\sum_{k=1}^{\infty}\sum_{p=1}^{\infty} \frac{p^{\frac{n+4}{2}}}{(ka)^{\frac{n}{2}}}K_{\frac{n}{2}}(4\pi kp aT)\right\},
\end{split}
\end{equation}
or
\begin{equation}\label{eq2_11_8}
\begin{split}
F_{\text{Cas}}^{\parallel} (a)\sim (n+2) \mathcal{V}(\mathcal{S})&\left\{- \frac{(n+3)\Gamma\left(\frac{n+4}{2}\right)}{(4\pi)^{\frac{n+4}{2}}a^{n+4}}\zeta_R(n+4)
-\frac{\Gamma\left(\frac{n+4}{2}\right)}{\pi^{\frac{n+4}{2}}}\zeta_R(n+4)T^{n+4}+\frac{\pi T^{\frac{n+1}{2}}}{2^{\frac{n+1}{2}}a^{\frac{n+7}{2}}}\sum_{k=1}^{\infty}\sum_{p=1}^{\infty}\frac{k^{\frac{n+5}{2}}}{p^{\frac{n+1}{2}}}K_{\frac{n+1}{2}}
\left(\frac{\pi kp}{aT}\right)\right\}
\\
\mp\frac{ n \mathcal{V}(\pa\mathcal{S})}{4}&\left\{- \frac{(n+2)\Gamma\left(\frac{n+3}{2}\right)}{(4\pi)^{\frac{n+3}{2}}a^{n+3}}\zeta_R(n+3)
-\frac{\Gamma\left(\frac{n+3}{2}\right)}{\pi^{\frac{n+3}{2}}}\zeta_R(n+3)T^{n+3}+\frac{\pi T^{\frac{n}{2}}}{2^{\frac{n}{2}}a^{\frac{n+6}{2}}}\sum_{k=1}^{\infty}\sum_{p=1}^{\infty}\frac{k^{\frac{n+4}{2}}}{p^{\frac{n}{2}}}K_{\frac{n}{2}}
\left(\frac{\pi pk}{aT}\right)\right\}.
\end{split}
\end{equation} Here $\mathcal{S}=\Omega\times\mathcal{N}$ is the cross section of the cylinder, $\pa\mathcal{S}$ is the boundary of $\mathcal{S}$; $\mathcal{V}(\mathcal{S})$ and $\mathcal{V}(\pa\mathcal{S})$ are respectively the volumes of $\mathcal{S}$ and $\pa\mathcal{S}$. The first terms in \eqref{eq2_11_7} and \eqref{eq2_11_8} are the Casimir force acting on a pair of perfectly conducting plates in the $N=4+n$ dimensional Minkowski spacetime \cite{13}.
When $aT\gg 1$, the leading term of the Casimir force is
$$F_{\text{Cas}}^{\parallel} (a)\sim-   \mathcal{V}(\mathcal{S})  T\frac{(n+2)^2\Gamma\left(\frac{n+3}{2}\right)}{(4\pi)^{\frac{n+3}{2}}a^{n+3}}\zeta_R(n+3);$$
and when $aT\ll 1$, the leading term of the Casimir force is
$$F_{\text{Cas}}^{\parallel} (a)\sim-  \mathcal{V}(\mathcal{S})  \frac{(n+2)(n+3)\Gamma\left(\frac{n+4}{2}\right)}{(4\pi)^{\frac{n+4}{2}}a^{n+4}}\zeta_R(n+4).$$
In the high temperature regime, we find that   the leading term of the Casimir force is of order $T/a^{n+3}$ if the separation between the plates is much smaller than the sizes of $\Omega$ and $\mathcal{N}$, but is of order $T/a^3$ if the size of $\mathcal{N}$ is much smaller than the separation between the plates, and the separation between the plates is much smaller than the size of $\Omega$. In the low temperature regime, we find that   the leading term of the Casimir force is of order $1/a^{n+4}$ if the separation between the plates is much smaller than the sizes of $\Omega$ and $\mathcal{N}$, but is of order $1/a^4$ if the size of $\mathcal{N}$ is much smaller than the separation between the plates, and the separation between the plates is much smaller than the size of $\Omega$. Therefore, we find that the strength of the Casimir force depends strongly on the relative magnitude of $a, r$ and $R$.

\subsection{The cylinder and the piston are imposed with different boundary conditions }\label{s4_2} If the cylinder is perfectly conducting and the piston is infinitely permeable,  the piston divides the cylinder $[0,L]\times\Omega\times\mathcal{N}$ into two cylinders $[0,a]\times \Omega\times\mathcal{N}$ and $[0,L-a]\times \Omega\times\mathcal{N}$, both of them have perfectly conducting sidewall and bottom, and infinitely permeable top. If the cylinder is infinitely permeable and the piston is perfectly conducting,  then the to cylinders have infinitely permeable sidewall and bottom, and perfectly conducting top. From the results of Section \ref{s3_3} and Section \ref{s3_4}, we find that  the zeta function
$\zeta_T(s;a)$ can be written as
\begin{equation*}\begin{split}
\zeta_T(s;a)=&  \sum_{p=-\infty}^{\infty}\sum_{\alpha}\zeta_{\alpha,p}(s;a)+\mathcal{C}(s),\\
\zeta_{\alpha,p}(s;a)=&\sum_{k=0}^{\infty}\left(\left[\frac{\pi\left(k+\frac{1}{2}\right)}{a}\right]^2+\tau_{\alpha,p}^2\right)^{-s},
\end{split}\end{equation*}where the set of $\tau_{\alpha}^2$ is given by (PC) if the cylinder is perfectly conducting and (IP) if the cylinder is infinitely permeable.

Using the fact that
\begin{equation*}
\sum_{k=0}^{\infty}\exp\left(-t\left[\frac{\pi \left(k+\frac{1}{2}\right)}{a}\right]^2\right)=\frac{a}{2\sqrt{\pi}}t^{-\frac{1}{2}}+ \frac{a}{\sqrt{\pi}}t^{-\frac{1}{2}}\sum_{k=1}^{\infty} (-1)^k \exp\left(-\frac{k^2a^2}{t}\right),
\end{equation*}we find as in Section \ref{s4_1} that if $\tau_{\alpha,p}^2\neq 0$,
\begin{equation*}
\begin{split}
\zeta_{\alpha,p}(0;a)=&\mathcal{C}_{\alpha,p}(0)+a\mathcal{D}_{\alpha,p}(0),\\
\zeta_{\alpha,p}'(0;a)
=&\mathcal{C}_{\alpha,p}'(0)+a\mathcal{D}_{\alpha,p}'(0)+  \sum_{k=1}^{\infty} \frac{(-1)^k}{k}e^{-2ka\tau_{\alpha,p}}.
\end{split}
\end{equation*}On the other hand, if $\tau_{\alpha,p}^2=0$,
\begin{align*}
\zeta_{\alpha,p}(s)=(2^{2s}-1)\left(\frac{\pi}{a}\right)^{-2s}\zeta_R(2s).
\end{align*}Therefore,
\begin{equation*}
\zeta_{\alpha,p}(0;a)=0\hspace{1cm}\text{and}\hspace{1cm}\zeta_{\alpha,p}'(0;a)=-\log 2.
\end{equation*}Hence, the Casimir energy of the piston system is
\begin{equation*}\begin{split}
E_{\text{Cas}}^{\text{piston}}=&\mathcal{E}_0-\frac{T}{2} \sum_{k=1}^{\infty}\sum_{\tau_{\alpha,p}\neq 0}\frac{(-1)^k}{k}e^{-2ka\tau_{\alpha,p}}-\frac{T}{2} \sum_{k=1}^{\infty}\sum_{\tau_{\alpha,p}\neq 0}\frac{(-1)^k}{k}e^{-2k(L-a)\tau_{\alpha,p}},
\end{split}
\end{equation*}where $\mathcal{E}_0$ is independent of $a$. It follows that the Casimir force acting on the piston is given by \eqref{eq1_25_2}, where the Casimir force acting between a pair of parallel plates with different boundary conditions inside the infinitely long cylinder is
\begin{equation}\label{eq1_25_6}\begin{split}
F_{\text{Cas}}^{\parallel}(a)=& -\frac{\pa  }{\pa a}\left(-\frac{T}{2} \sum_{k=1}^{\infty}\sum_{\tau_{\alpha,p}\neq 0}\frac{(-1)^k}{k}e^{-2ka\tau_{\alpha,p}}\right)
\\=&-T \sum_{k=1}^{\infty}\sum_{\tau_{\alpha,p}}(-1)^k\tau_{\alpha,p}e^{-2ka\tau_{\alpha,p}} =
T \sum_{ \tau_{\alpha,p}}\frac{\tau_{\alpha,p}}{e^{2a\tau_{\alpha,p}}+1}.
\end{split}\end{equation}Notice that contrary to the previous case where the two plates are both infinitely permeable, at finite temperature the force acting on a pair of plates, one perfectly conducting and one infinitely permeable, does not have a long range term even though there are TEM modes for which $\tau_{\alpha,p}^2=0$.

 Eq. \eqref{eq1_25_6} shows that the Casimir force acting between one perfectly conducting plate and one infinitely permeable plate is always repulsive, and is a monotonically decreasing function of the distance between the plates. Therefore, in a closed   cylinder with finite length, the Casimir force acting on the piston which have  different boundary conditions  always tends to push the   piston to the middle of the cylinder, which is the equilibrium position.

In the following, we study the asymptotic behavior of the Casimir force  \eqref{eq1_25_6} at different limits. In the high temperature limit $aT\gg 1$, the Casimir force \eqref{eq1_25_6} is dominated by the classical term given by
\begin{equation*} \begin{split}
F_{\text{Cas}}^{\parallel}(a)\sim
T \sum_{ \tau_{\alpha}}\frac{\tau_{\alpha,p}}{e^{2a\tau_{\alpha}}+1}.
\end{split}\end{equation*}The remaining terms decay exponentially. The result of Section \ref{a_2} shows that the Casimir force \eqref{eq1_25_6} can also be written as
\begin{equation*}
\begin{split}
F_{\text{Cas}}^{\parallel}(a)
=&\sum_{\tau_{\alpha}=0}\left\{\frac{\pi}{48a^2}-\frac{\pi T^2}{6} \right\}+\sum_{\tau_{\alpha}\neq 0}\left\{\frac{1}{2\pi a}\sum_{k=1}^{\infty}(-1)^{k-1}\frac{\tau_{\alpha}}{k} K_1(2ka\tau_{\alpha})+\frac{1}{\pi}\sum_{k=1}^{\infty}(-1)^{k-1} \tau_{\alpha}^2K_0(2ka\tau_{\alpha})
-\frac{T}{\pi}\sum_{p=1}^{\infty}\frac{\tau_{\alpha}}{p}K_1\left(\frac{p\tau_{\alpha}}{T}\right)\right\}\\&+\frac{\pi^2}{a^3}
\sum_{\tau_{\alpha}}\sum_{k=0}^{\infty}
\frac{\left(k+\frac{1}{2}\right)^2}{\sqrt{\left[\frac{\pi \left(k+\frac{1}{2}\right)}{a}\right]^2+\tau_{\alpha}^2}}\left\{\exp\left(\frac{1}{T}\sqrt{\left[\frac{\pi \left(k+\frac{1}{2}\right)}{a}\right]^2+\tau_{\alpha}^2} \right) -1\right\}^{-1}.
\end{split}
\end{equation*}Therefore the zero temperature Casimir force is given by
\begin{equation*}
\begin{split}
F_{\text{Cas}}^{\parallel, T=0}(a)
=&\sum_{\tau_{\alpha}=0} \frac{\pi}{48a^2} +\sum_{\tau_{\alpha}\neq 0}\left\{\frac{1}{2\pi a}\sum_{k=1}^{\infty}(-1)^{k-1}\frac{\tau_{\alpha}}{k} K_1(2ka\tau_{\alpha})+\frac{1}{\pi}\sum_{k=1}^{\infty}(-1)^{k-1} \tau_{\alpha}^2K_0(2ka\tau_{\alpha})
 \right\}
\end{split}
\end{equation*}Notice that the zero temperature Casimir force contains a long range term if the plates are placed inside an infinitely permeable cylinder and the first Betti number $h$ of $\mathcal{N}$ is nonzero. In this case, the leading term of the thermal correction is also of order $T^2$.

Using the results of Appendix \ref{b}, we can also derive the leading terms of the Casimir force when $r\ll a\ll R$ and $a\ll r, R$. In the high temperature regime, we find that if $r\ll a\ll R$, then
\begin{equation}
\begin{split}
F_{\text{Cas}}^{\parallel} (a)\sim &  \frac{2+h}{2} \mathcal{A}(\Omega) \frac{3T }{16\pi a^3}\zeta_R(3),
 \end{split}
\end{equation}and if $a\ll r,R$,
\begin{equation}
\begin{split}
F_{\text{Cas}}^{\parallel} (a)\sim &  \mathcal{V}(\mathcal{S})  T\frac{(n+2)^2\Gamma\left(\frac{n+3}{2}\right)}{(4\pi)^{\frac{n+3}{2}}a^{n+3}}\zeta_R(n+3)\left(1-2^{-n-2}\right).
 \end{split}
\end{equation}
In the low temperature regime, if $r\ll a\ll R$,
\begin{equation}
\begin{split}
F_{\text{Cas}}^{\parallel} (a)\sim &  \frac{2+h}{2}\mathcal{A}(\Omega)   \frac{7\pi^2}{1920a^4},
 \end{split}
\end{equation} and if $a\ll r, R$,
$$F_{\text{Cas}}^{\parallel} (a)\sim   \mathcal{V}(\mathcal{S})  \frac{(n+2)(n+3)\Gamma\left(\frac{n+4}{2}\right)}{(4\pi)^{\frac{n+4}{2}}a^{n+4}}\zeta_R(n+4)\left(1-2^{-n-3}\right).$$
Again we find that in the high temperature regime,    the leading term of the Casimir force is of order $T/a^{n+3}$ when $a\ll r, R$, but is of order $T/a^3$ when $r\ll a\ll R$. In the low temperature regime,    the leading term of the Casimir force is of order $1/a^{n+4}$ when $a\ll r, R$, but is of order $1/a^4$ when $r\ll a\ll R$. We also find that when the size of the manifold $\mathcal{N}$ goes to zero, then the Casimir force reduces to the Casimir force in the $4D$ Minkowski spacetime if and only if the first Betti number $h$ of $\mathcal{N}$ is zero. Otherwise, there are some extra contributions.

\section{Explicit examples}\label{s5}
In this section, we consider the specific examples where the manifold $\mathcal{N}$ is an $n $-dimensional torus $T^n$ or an $n $-dimensional sphere $S^n$ with volume $r^n$. Assume that $T^n$ are $n$ copies of $S^1$ with the same radius $r_1$. Then the volume of $T^n$ is $r^n$ implies that $r_1=r/(2\pi)$. The radius of the sphere $S^n$ with volume $r^n$ is
$$r_2= \left(\frac{\Gamma\left(\frac{n+1}{2}\right)}{2\pi^{\frac{n+1}{2}}}\right)^{\frac{1}{n}}r.$$

On $T^n$, the spectrum $\{m_j\,:\,j=0,1,2,\ldots\}$ of the Laplace operator on functions is given by
\begin{equation}\label{eq2_16_1}\left\{\frac{  j_1^2+\ldots+j_n^2}{r_1^2}\,:\,j_1,\ldots,j_n\in \mathbb{Z} \right\},\end{equation}
and the spectrum $ \{\mu_j^2\,:\,j=1,2,\ldots\}$ of the Laplace operator on co-closed one-forms is given by $(n-1)$ copies of \eqref{eq2_16_1} plus one zero.   $h=n$ for $T^n$.

On $S^n$, the spectrum $\{m_j\,:\,j=0,1,2,\ldots\}$ of the Laplace operator on functions is given by \cite{33}:
\begin{equation*} b_l^0= \frac{l(l+n-1)}{r_1^2},\quad l=0,1,2,\ldots,\end{equation*} with multiplicities
$$d_l^0=\frac{(2l+n-1)(l+n-2)!}{l!(n-1)!};$$
and the spectrum $ \{\mu_j^2\,:\,j=1,2,\ldots\}$ of the Laplace operator on co-closed one-forms is given by \cite{33}:
\begin{equation}
b_l^1=l(l+n-1)+n-2,\quad l=1,2,\ldots,
\end{equation}with multiplicities
$$d_l^1=\frac{l(l+n-1)(2l+n-1)(l+n-3)!}{(n-2)!(l+1)!}.$$  $h=0$ for $S^n$.

We only consider the case where the size of the domain $\Omega$ is much larger than $a$ and $r$. Then the Casimir force is given by the first term in \eqref{eq2_11_1} or
\eqref{eq2_11_2}.
\begin{figure}[h]\centering
\epsfxsize=0.49\linewidth \epsffile{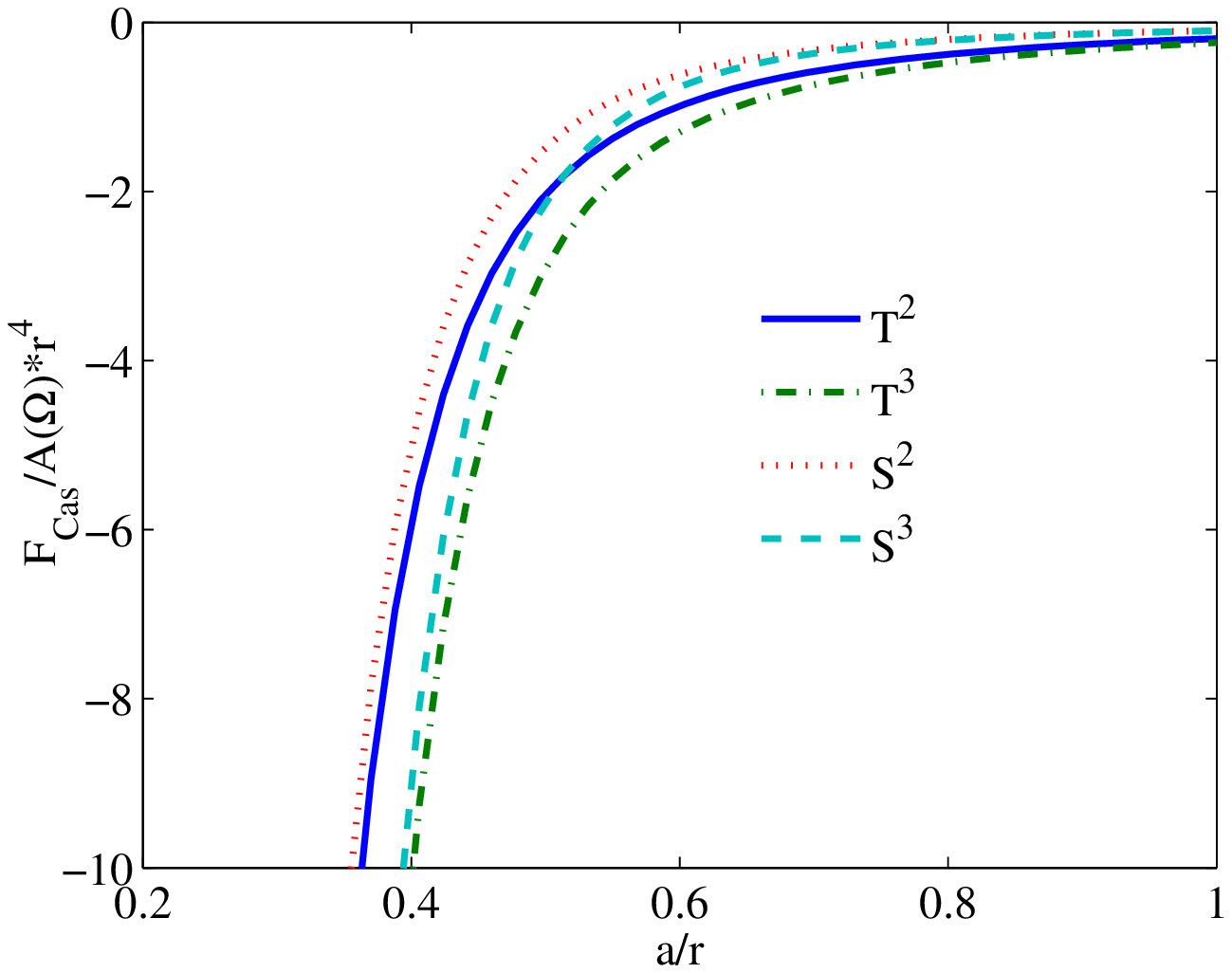}\centering
\epsfxsize=0.49\linewidth \epsffile{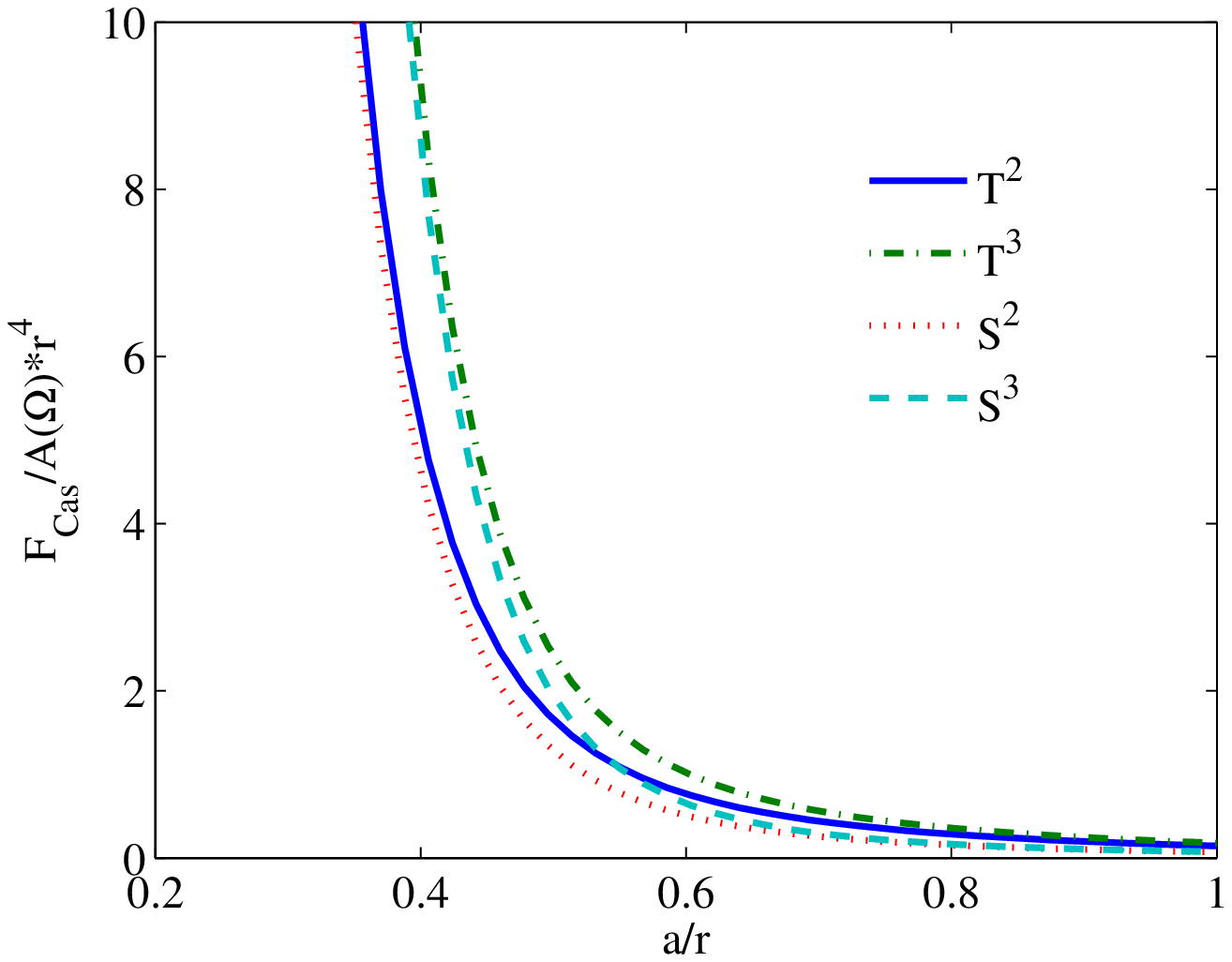} \caption{\label{f4} $\displaystyle \frac{F_{\text{Cas}}^{\parallel}}{A(\Omega)}r^4$ as a function of $a/r$ for $\mathcal{N}=T^2, T^3, S^2, S^3$. Here $rT=1$. For the graph on the left, the two plates are both perfectly conducting or infinitely permeable. For the graph on the right, one plate is perfectly conducting and one is infinitely permeable.}\end{figure}

\begin{figure}[h]\centering
\epsfxsize=0.49\linewidth \epsffile{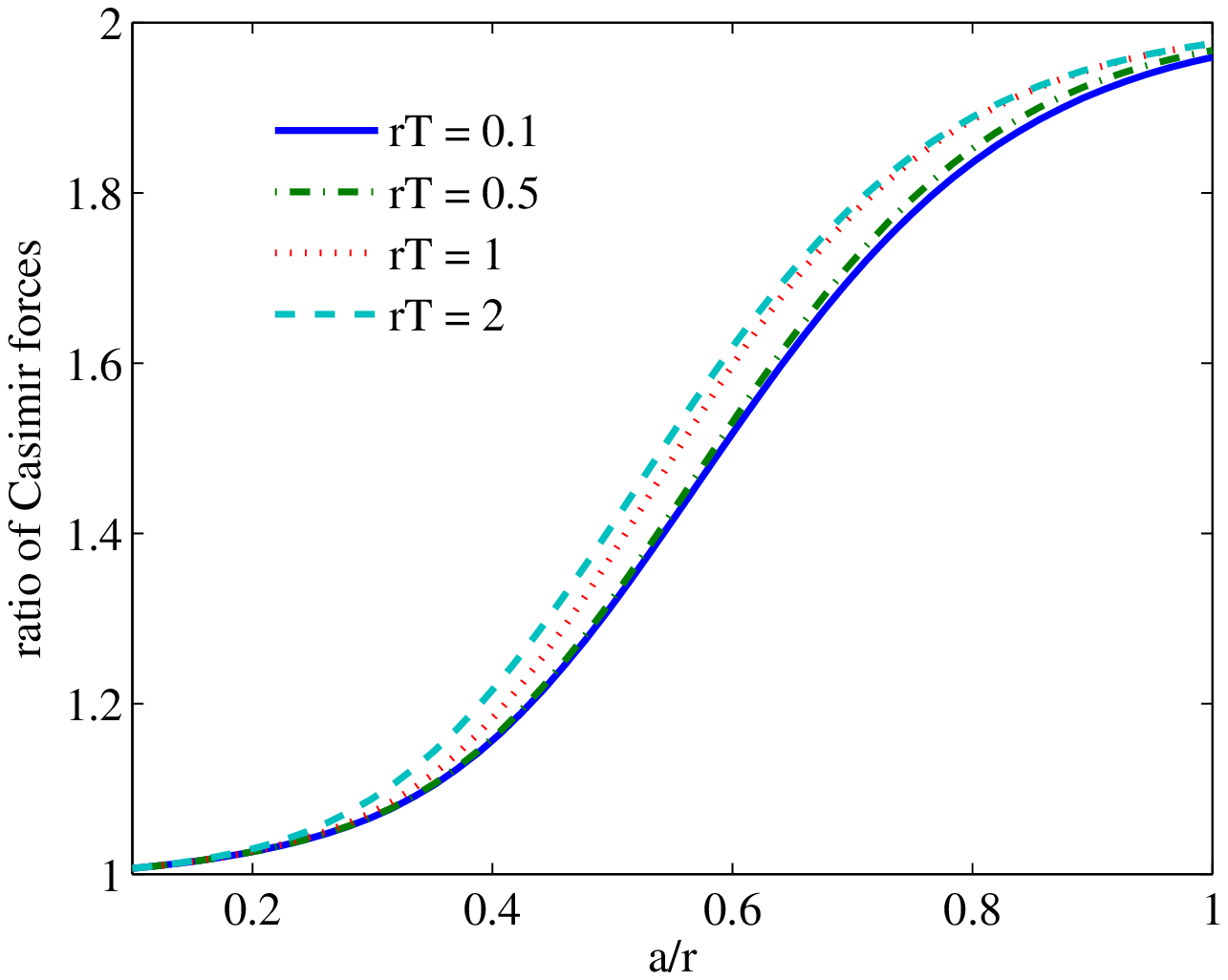}\centering
\epsfxsize=0.49\linewidth \epsffile{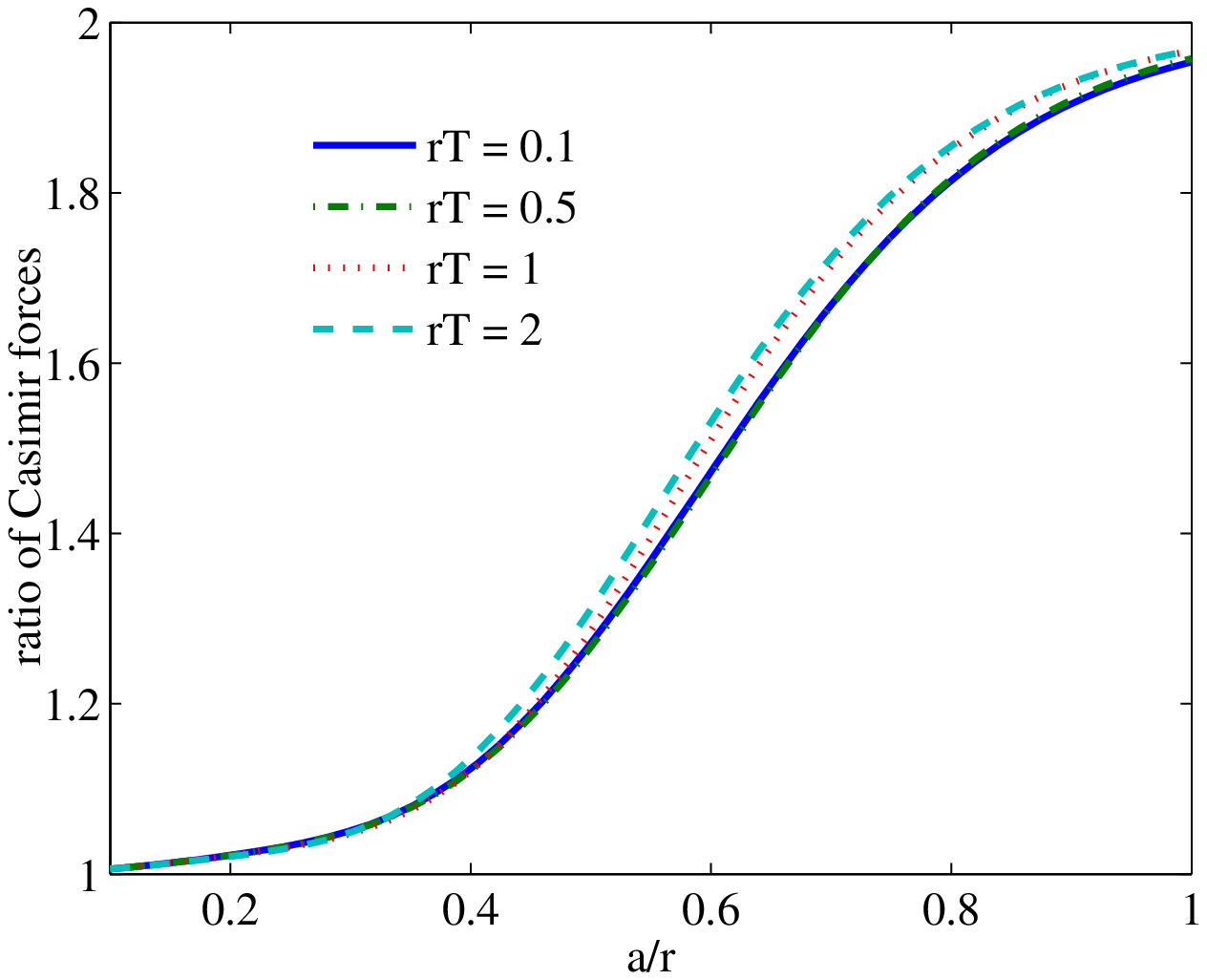} \caption{\label{f2} The ratio of the Casimir force on $T^2$ to the Casimir force on $S^2$ as a function of $a/r$, for $rT=0.1,0.5,1,2$. For the graph on the left, the two plates are both perfectly conducting or infinitely permeable. For the graph on the right, one plate is perfectly conducting and one is infinitely permeable.}\end{figure}

\begin{figure}[h]\centering
\epsfxsize=0.49\linewidth \epsffile{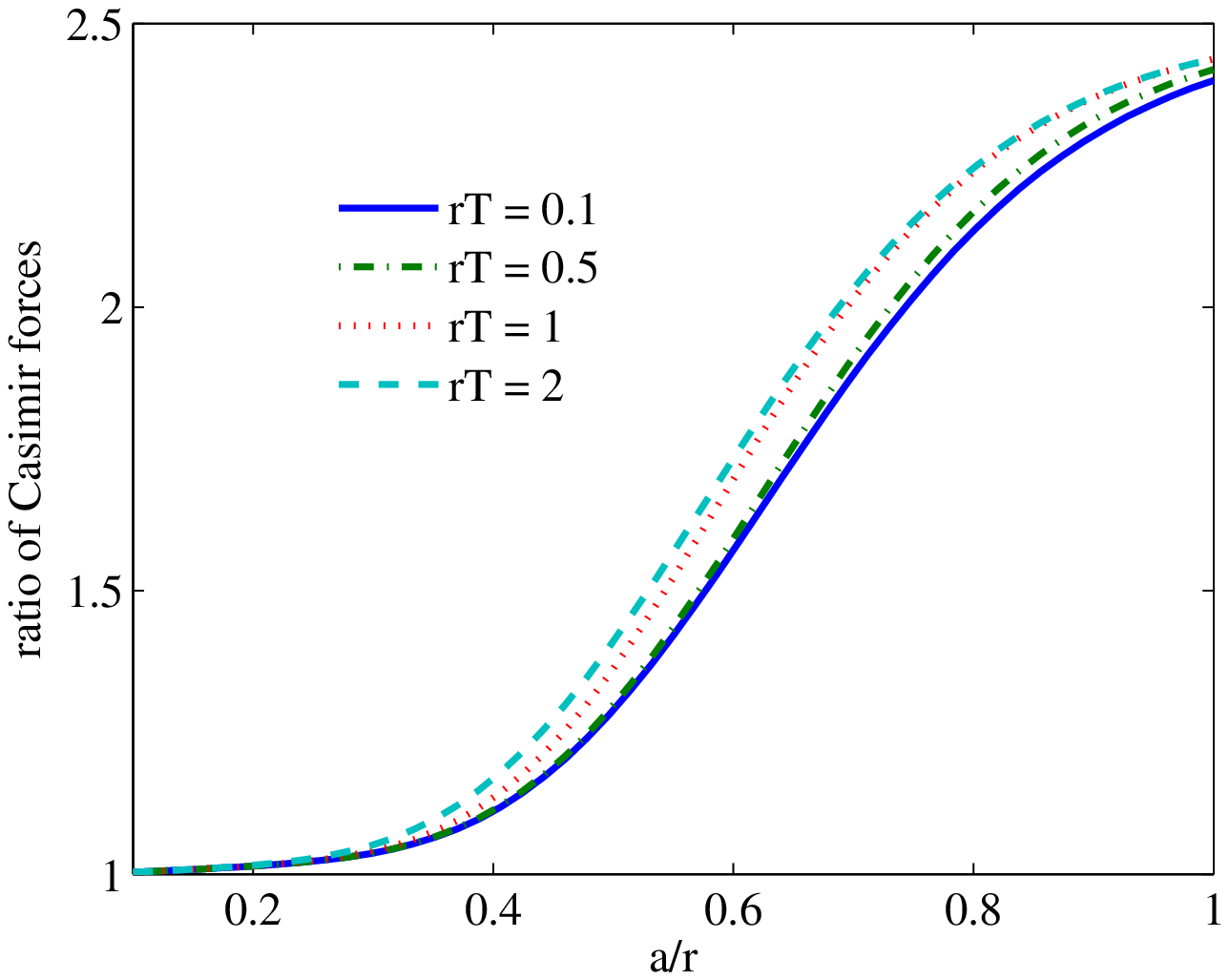}\centering
\epsfxsize=0.49\linewidth \epsffile{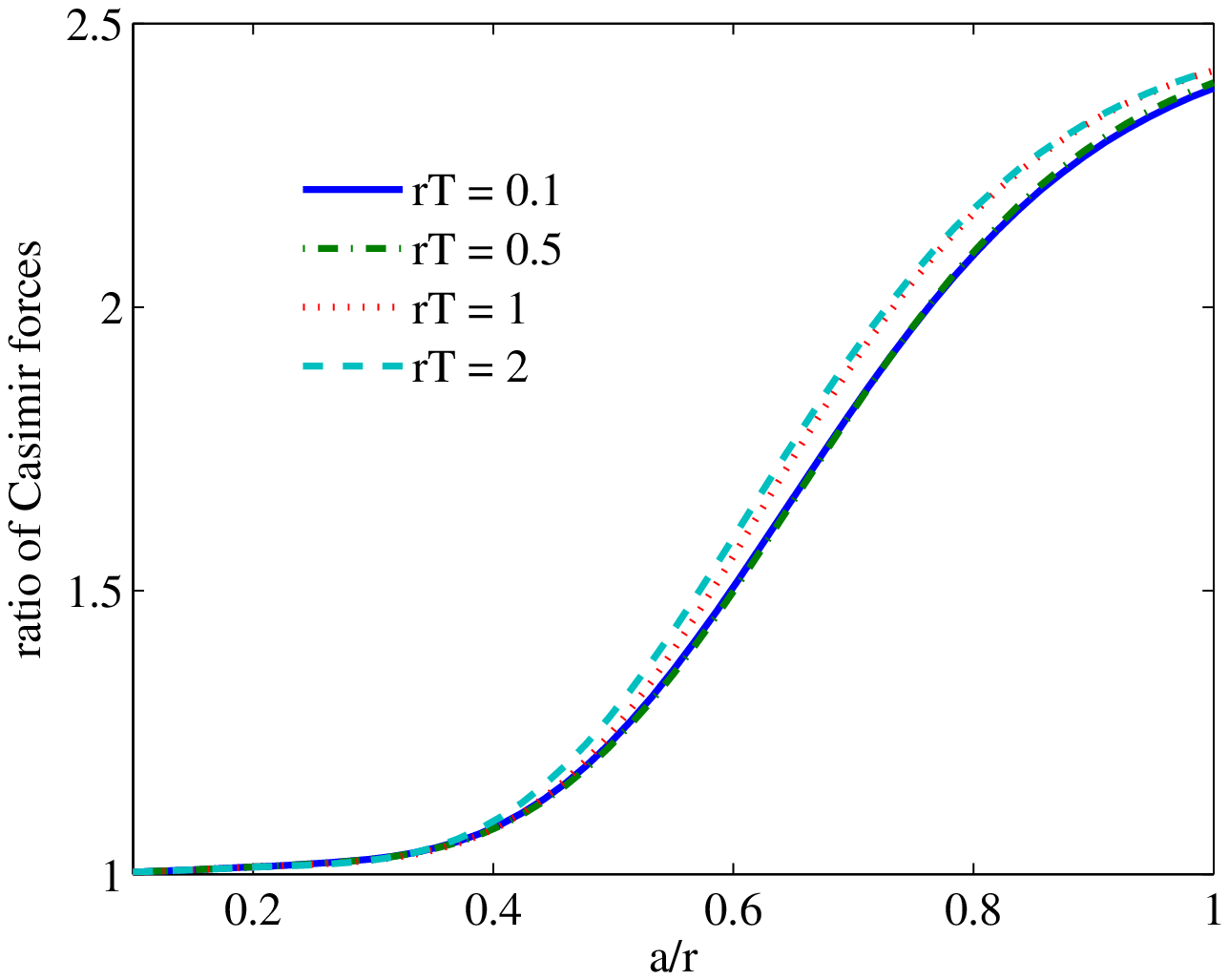} \caption{\label{f3} The ratio of the Casimir force on $T^3$ to the Casimir force on $S^3$ as a function of $a/r$, for $rT=0.1,0.5,1,2$. For the graph on the left, the two plates are both perfectly conducting or infinitely permeable. For the graph on the right, one plate is perfectly conducting and one is infinitely permeable.}\end{figure}

In Figure \ref{f4}, we show the graphs of the dimensionless Casimir force $\displaystyle \frac{F_{\text{Cas}}^{\parallel}}{A(\Omega)}r^4$ as a function of $a/r$ when $rT=1$ for $\mathcal{N}=T^2,T^3,S^2$ and $S^3$. From these graphs, one can see that as $a/r$ gets smaller,  the magnitude of the Casimir force for the same  value of $n$ agrees more, and it is larger for larger $n$, in agreement with the $(rT)/(a/r)^{n+3}$ or $1/(a/r)^{n+4}$ behavior for small $a/r$. It is not easy to read from the  graphs in Figure \ref{f4} the difference between the Casimir forces when $a/r$ gets larger. Therefore, in Figure \ref{f2} and Figure \ref{f3}, we show the ratio of the Casimir force when $\mathcal{N}$ is $T^n$ to the Casimir force when $\mathcal{N}$ is $S^n$. We plot the graphs for $rT=0.1, 0.5, 1$ and $2$. The $n=2$ case is shown in Figure \ref{f2} and the $n=3$ case is shown in Figure \ref{f3}. These graphs show that for small $a/r$, the ratios of the Casimir forces for fixed $n$ indeed approaches unity. For larger $a/r$, the ratios of the Casimir forces approaches $1+n/2$, in agreement with the fact that as $a\gg r$, the leading term of the Casimir force is $1+h/2$ times the Casimir force in the $4D$ Minkowski spacetime. Since $h=n$ for $T^n$ and $h=0$ for $S^n$, therefore the ratio of the Casimir force    when $\mathcal{N}$ is $T^n$ to the Casimir force when $\mathcal{N}$ is $S^n$ should approach $1+n/2$ when $a\gg r$.

\section{Conclusion}
In this article, we have investigated the electromagnetic Casimir effect on a piston in a higher dimensional spacetime of the form $M\times\mathcal{N}$. One of the reasons the Casimir effect of electromagnetic fields is a much more difficult problem  than the Casimir effect of scalar fields is that electromagnetic fields have gauge degree of freedom. It is very important to choose gauges that eliminate all the gauge freedom and which  facilitate the correct counting of eigenmodes of the electromagnetic field.  We have discussed this issue in detail in Section \ref{BF} and Section \ref{s3}.

In Section \ref{s4}, we compute the Casimir force acting on the piston for different combinations of perfectly conducting boundary conditions and infinitely permeable boundary conditions on the cylinder and on the piston.    It is shown that  if the cylinder and the piston have the same boundary conditions, then the Casimir force always tends to pull the piston to the closer end of the cylinder. However, if the cylinder and the piston have different boundary conditions, the Casimir force always tends to push the piston to the equilibrium position in the middle of the cylinder. It is also discovered that if the cylinder is closed and infinitely permeable, then the Casimir force has a long range term if the first Betti number $h$ of $\mathcal{N}$ which counts the dimension  of harmonic one-forms on $\mathcal{N}$ is nonzero.

The asymptotic behavior of the Casimir force acting on a pair of parallel plates which are obtained by taking one end of the cylinder to infinity is discussed.  We consider the high temperature asymptotic behavior and  the low temperature asymptotic behavior for different relative magnitude of $a$ -- the separation between the plates, $r$ -- the size of $\mathcal{N}$, and $R$ -- the size of the cross-section of the cylinder in $M$. It is shown that if $a\ll r, R$, then the leading term of the Casimir force is the Casimir force between two large parallel plates in the $(4+n)$-dimensional Minkowski spacetime, which is of order $1/a^{n+4}$ in the low temperature regime, and of order $T/a^{n+3}$ in the high temperature regime. On the other hand, if $r\ll a\ll R$, then the leading term of the Casimir force is $1+h/2$ times the  Casimir force between two large parallel plates in the $4$-dimensional Minkowski spacetime. In particular, this shows that if the size $r$ of $\mathcal{N}$ reduces to zero, one will obtain the Casimir force in the $4$-dimensional Minkowski spacetime if and only if $h$, the number of zero modes of the Laplace operator on one-forms on $\mathcal{N}$, is zero. This poses a topological condition for the smooth transition of the Casimir force from the spacetime $M\times \mathcal{N}$ to the spacetime $M$.

\appendix
\section{Alternative expressions for the Casimir force}
\subsection{The plates have  the same boundary conditions }\label{a_1}
As shown in Section \ref{s4_1}, the Casimir force  acting between two plates with the same boundary conditions is the sum over all $\tau_{\alpha}$ of  the following expression:
$$-T\sum_{p=-\infty}^{\infty}\frac{\tau_{\alpha,p}}{e^{2a\tau_{\alpha,p}}-1}. $$
This expression can be written as
\begin{equation*}
\begin{split}
&\frac{T}{2}\frac{\pa}{\pa a} \sum_{p=-\infty}^{\infty}\sum_{k=1}^{\infty}\frac{1}{k}e^{-2ka\sqrt{\tau_{\alpha}^2+[2\pi p T]^2}}\\
=&\frac{T}{2\sqrt{\pi}}\frac{\pa}{\pa a} \left\{a\int_0^{\infty}t^{-\frac{1}{2}}\sum_{p=-\infty}^{\infty}\sum_{k=1}^{\infty}\exp\left(-tk^2a^2-\frac{\tau_{\alpha}^2+[2\pi p T]^2}{t}\right)dt\right\}\\
=&\frac{1}{4\pi}\frac{\pa}{\pa a} \left\{a\int_0^{\infty} \sum_{p=-\infty}^{\infty}\sum_{k=1}^{\infty}\exp\left(-tk^2a^2-\frac{tp^2}{4T^2}-\frac{\tau_{\alpha}^2 }{t}\right)dt\right\}\\
=&\frac{1}{4\pi}\frac{\pa}{\pa a} \left\{a\int_0^{\infty}  \sum_{k=1}^{\infty}\exp\left( -tk^2a^2-\frac{\tau_{\alpha}^2 }{t}\right)dt
-a\int_0^{\infty}  \sum_{p=1}^{\infty}\exp\left( -\frac{tp^2}{4T^2}-\frac{\tau_{\alpha}^2 }{t}\right)dt
\right.\\&\left.+2\sqrt{\pi}\int_0^{\infty} t^{-\frac{1}{2}}\sum_{p=1}^{\infty}\sum_{k=0}^{\infty}\!'\exp\left(- \frac{tp^2}{4T^2}-\frac{1 }{t}\left(\left[\frac{\pi k}{a}\right]^2+\tau_{\alpha}^2\right)\right)dt\right\}.\end{split}
\end{equation*}
If $\tau_{\alpha}^2\neq 0$, this is equal to
\begin{equation*}
\begin{split}
&\frac{1}{4\pi}\frac{\pa}{\pa a} \left\{2 \sum_{k=1}^{\infty}\frac{\tau_{\alpha}}{k}K_1(2ka\tau_{\alpha})-4aT\sum_{p=1}^{\infty}\frac{\tau_{\alpha}}{p}K_1\left(\frac{p\tau_{\alpha}}{T}\right)+4\pi T\sum_{p=1}^{\infty}\sum_{k=1}^{\infty}
\frac{1}{p}\exp\left(-\frac{p\sqrt{\left[\frac{\pi k}{a}\right]^2+\tau_{\alpha}^2}}{T}\right)\right\}\\
=&-\frac{1}{2\pi a}\sum_{k=1}^{\infty}\frac{\tau_{\alpha}}{k} K_1(2ka\tau_{\alpha})-\frac{1}{\pi}\sum_{k=1}^{\infty} \tau_{\alpha}^2K_0(2ka\tau_{\alpha})
-\frac{T}{\pi}\sum_{p=1}^{\infty}\frac{\tau_{\alpha}}{p}K_1\left(\frac{p\tau_{\alpha}}{T}\right)+\frac{\pi^2}{a^3}
\sum_{k=1}^{\infty}
\frac{k^2}{\sqrt{\left[\frac{\pi k}{a}\right]^2+\tau_{\alpha}^2}}\frac{1}{\exp\left(\frac{1}{T}\sqrt{\left[\frac{\pi k}{a}\right]^2+\tau_{\alpha}^2} \right) -1}.
\end{split}
\end{equation*}
But if $\tau_{\alpha}^2=0$, it is equal to
\begin{equation*}
\begin{split}
&\frac{1}{4\pi}\frac{\pa}{\pa a} \left\{\frac{\pi^2}{6a}-\frac{2\pi^2}{3}aT^2 +4\pi T\sum_{p=1}^{\infty}\sum_{k=1}^{\infty}
\frac{1}{p}\exp\left(-\frac{\pi k p}{aT}\right)\right\}\\
=&- \frac{\pi}{24a^2}-\frac{\pi T^2}{6}+\frac{\pi}{a^2}
\sum_{k=1}^{\infty}
 \frac{k}{\exp\left(\frac{\pi k }{aT}\right) -1}.
\end{split}
\end{equation*}
\subsection{The plates have different boundary conditions }\label{a_2}

As shown in Section \ref{s4_2}, the Casimir force  acting between two plates with different boundary conditions is the sum over all  $\tau_{\alpha}$ of  the following expression:
$$T\sum_{p=-\infty}^{\infty}\frac{\tau_{\alpha,p}}{e^{2a\tau_{\alpha,p}}+1} =\frac{T}{2}\frac{\pa}{\pa a} \sum_{p=-\infty}^{\infty}\sum_{k=1}^{\infty}\frac{(-1)^k}{k}e^{-2ka\sqrt{\tau_{\alpha}^2+[2\pi p T]^2}}.$$
As in Section \ref{a_1}, one can show that if $\tau_{\alpha}^2\neq 0$, this expression  is equal to
\begin{equation*}
\begin{split}
 &-\frac{1}{2\pi a}\sum_{k=1}^{\infty}(-1)^k\frac{\tau_{\alpha}}{k} K_1(2ka\tau_{\alpha})-\frac{1}{\pi}\sum_{k=1}^{\infty}(-1)^k \tau_{\alpha}^2K_0(2ka\tau_{\alpha})
-\frac{T}{\pi}\sum_{p=1}^{\infty}\frac{\tau_{\alpha}}{p}K_1\left(\frac{p\tau_{\alpha}}{T}\right)\\&+\frac{\pi^2}{a^3}
\sum_{k=0}^{\infty}
\frac{\left(k+\frac{1}{2}\right)^2}{\sqrt{\left[\frac{\pi \left(k+\frac{1}{2}\right)}{a}\right]^2+\tau_{\alpha}^2}}\left\{\exp\left(\frac{1}{T}\sqrt{\left[\frac{\pi \left(k+\frac{1}{2}\right)}{a}\right]^2+\tau_{\alpha}^2} \right) -1\right\}^{-1},
\end{split}
\end{equation*} and if $\tau_{\alpha}^2=0$, it is equal to
 \begin{equation*}
\begin{split}
& \frac{\pi}{48a^2}-\frac{\pi T^2}{6}+\frac{\pi}{a^2}
\sum_{k=0}^{\infty}
 \frac{k+\frac{1}{2}}{\exp\left(\frac{\pi\left( k+\frac{1}{2}\right) }{aT}\right) -1}.
\end{split}
\end{equation*}

\section{Asymptotic behavior of the Casimir force}\label{b}
 To find the asymptotic behavior of the Casimir force when $r\ll a\ll R$ and $a\ll r, R$, let us define the global heat kernels
\begin{equation*}
\begin{split}
&K_{\Omega,D}(t)=\sum_{l=1}^{\infty}e^{-t\varpi_l^{2}},\hspace{1cm}K_{\Omega,N}(t)=\sum_{l=0}^{\infty}e^{-t\varkappa_l^2},\\
&K_{\mathcal{N},0}(t)=\sum_{j=0}^{\infty} e^{-tm_j^{2}},\hspace{1cm}K_{\mathcal{N},1 }(t)=\sum_{j=1}^{\infty} e^{-t m_j^2}+\sum_{j=1}^{\infty}e^{-t\mu_j^{2}}.\end{split}
\end{equation*} $K_{\Omega,D}(t)$ is  the heat kernel of the Laplace operator with Dirichlet boundary conditions on functions on $\Omega$.  $K_{\Omega,N}(t)$ is  the heat kernel of the Laplace operator with Neumann boundary conditions on functions on $\Omega$. $K_{\mathcal{N},0}(t)$ is  the heat kernel of the Laplace operator  on functions on $\mathcal{N}$. For $K_{\mathcal{N},1 }(t)$, notice that given a one-form $V$ which is an eigenvector of the Laplace operator   with eigenvalue $m^2\neq 0$, if it is not co-closed, i.e., if $\delta_{\mathcal{N}} V\neq 0$, then $\delta_{\mathcal{N}} V$ is a nonzero function on $\mathcal{N}$. Moreover,
$$\Delta_{\mathcal{N}}(\delta_{\mathcal{N}} V)=(\delta_{\mathcal{N}}d_{\mathcal{N}}+d_{\mathcal{N}}\delta_{\mathcal{N}}) \delta_{\mathcal{N}} V=\delta_{\mathcal{N}}d_{\mathcal{N}} \delta_{\mathcal{N}}V=\delta_{\mathcal{N}}(d_{\mathcal{N}}\delta_{\mathcal{N}}+\delta_{\mathcal{N}} d_{\mathcal{N}})V=\delta_{\mathcal{N}} \Delta_{\mathcal{N}}V=m^2\delta_{\mathcal{N}} V.$$ Namely, $\delta_{\mathcal{N}} V$ is an eigenfunction of the Laplace operator   with eigenvalue $m^2$. Conversely, if $q$ is  an eigenfunction of the Laplace operator with eigenvalue $m^2\neq 0$, then $dq$ is a one-form, and
$$\Delta_{\mathcal{N}} (d_{\mathcal{N}}q)= (d_{\mathcal{N}}\delta_{\mathcal{N}}+\delta_{\mathcal{N}} d_{\mathcal{N}}) d_{\mathcal{N}}q=d_{\mathcal{N}}\delta_{\mathcal{N}} d_{\mathcal{N}}q= d_{\mathcal{N}}(\delta_{\mathcal{N}} d_{\mathcal{N}}+ d_{\mathcal{N}}\delta_{\mathcal{N}})q=d_{\mathcal{N}}(\Delta_{\mathcal{N}} q)=m^2d_{\mathcal{N}}q.$$Namely, $d_{\mathcal{N}}q$ is an eigen-one-form of the Laplace operator with eigenvalue $m^2$.
Therefore the union of $m_j^2, j\geq 1$, and $\mu_j^2, j\geq 1$, are the set of all eigenvalues of the Laplace operator on one-forms on $\mathcal{N}$. Hence, $K_{\mathcal{N},1}(t)$ is  the heat kernel of the Laplace operator  on one-forms on $\mathcal{N}$.

It is well known \cite{10,11,12} that as $t\rightarrow 0^+$, the heat kernels $K(t)$ have   asymptotic expansions of the form
$$K(t)\sim \sum_{i=0}^{\infty}c_i t^{\frac{i-d}{2}}=c_0t^{-\frac{d}{2}}+c_1t^{-\frac{d-1}{2}}+\ldots,$$where $d$ is the dimension of the manifold. More specifically,
\begin{equation*}
\begin{split}
K_{\Omega,D}(t)=&\frac{\mathcal{A}(\Omega)}{4\pi}t^{-1}-\frac{l(\pa\Omega)}{8\sqrt{\pi}}t^{-\frac{1}{2}}+O(1),\\
K_{\Omega,N}(t)=&\frac{\mathcal{A}(\Omega)}{4\pi}t^{-1}+\frac{l(\pa\Omega)}{8\sqrt{\pi}}t^{-\frac{1}{2}}+O(1),\\
K_{\mathcal{N},0}(t)=&\frac{\mathcal{V}(\mathcal{N})}{(4\pi)^{\frac{n}{2}}}t^{-\frac{n}{2}}+O\left(t^{1-\frac{n}{2}}\right),\\
K_{\mathcal{N},1}(t)=&n\frac{\mathcal{V}(\mathcal{N})}{(4\pi)^{\frac{n}{2}}}t^{-\frac{n}{2}}+O\left(t^{1-\frac{n}{2}}\right).
\end{split}
\end{equation*}Here $\mathcal{A}(\Omega)$ is the area of $\Omega$,  $l(\pa\Omega)$ is the arc length of $\pa\Omega$ and  $\mathcal{V}(\mathcal{N})$ is the volume of $\mathcal{N}$.

As is shown in Section \ref{a_1} and Section \ref{a_2}, the Casimir force acting on a pair of parallel plates is given by
\begin{equation}\label{eq1_27_1}
\begin{split}
F_{\text{Cas}}^{\parallel}(a)= \frac{T}{2\sqrt{\pi}}\frac{\pa}{\pa a} \left\{a\int_0^{\infty}t^{-\frac{1}{2}}\sum_{p=-\infty}^{\infty}\sum_{k=1}^{\infty}e^{2\pi i k\chi}\exp\left(-tk^2a^2-\frac{ [2\pi p T]^2}{t}\right)
\sum_{\tau_{\alpha}}\exp\left(-\frac{\tau_{\alpha}^2}{t}\right)dt\right\},
 \end{split}
\end{equation}or
\begin{equation}\label{eq1_27_2}
\begin{split}
F_{\text{Cas}}^{\parallel}(a)= &\frac{1}{4\pi}\frac{\pa}{\pa a} \left\{a\int_0^{\infty}  \sum_{k=1}^{\infty}e^{2\pi i k\chi}\exp\left( -tk^2a^2 \right)\sum_{\tau_{\alpha}}\exp\left(-\frac{\tau_{\alpha}^2}{t}\right)dt
-a\int_0^{\infty}  \sum_{p=1}^{\infty}\exp\left( -\frac{tp^2}{4T^2} \right)\sum_{\tau_{\alpha}}\exp\left(-\frac{\tau_{\alpha}^2}{t}\right)dt
\right.\\&\left.+2\sqrt{\pi}\int_0^{\infty} t^{-\frac{1}{2}}\sum_{p=1}^{\infty}\sum_{k=0}^{\infty} \exp\left(- \frac{tp^2}{4T^2}-\frac{1 }{t} \left(\frac{\pi (k+\chi)}{a}\right)^2 \right)\sum_{\tau_{\alpha}}\exp\left(-\frac{\tau_{\alpha}^2}{t}\right)dt\right\},\end{split}
\end{equation}
where $\chi=1$ if the two plates have the same boundary condition, and $\chi=1/2$ if the two plates have different boundary conditions.

When the cylinder is perfectly conducting, the set of $\tau_{\alpha}^2$ is given by (PC). Therefore,
\begin{equation}\label{eq2_8_1}
\begin{split}
\sum_{\tau_{\alpha}}\exp\left(-\frac{\tau_{\alpha}^2}{t}\right)=&2K_{\Omega,D}\left(t^{-1}\right)\sum_{j=0}^{\infty}\!'\exp\left(-\frac{m_j^2}{t}\right)
+K_{\Omega,D}\left(t^{-1}\right)\sum_{j=1}^{\infty}\exp\left(-\frac{\mu_j^2}{t}\right)+\left(K_{\Omega,N}\left(t^{-1}\right)-1\right)\sum_{j=0}^{\infty}
\exp\left(-\frac{m_j^2}{t}\right)\\
\sim &2\left(\frac{\mathcal{A}(\Omega)}{4\pi}t -\frac{l(\pa\Omega)}{8\sqrt{\pi}}t^{\frac{1}{2}}\right)\sum_{j=0}^{\infty}\!'\exp\left(-\frac{m_j^2}{t}\right)+\left(\frac{\mathcal{A}(\Omega)}{4\pi}t
-\frac{l(\pa\Omega)}{8\sqrt{\pi}}t^{\frac{1}{2}}\right) \sum_{j=1}^{\infty}\exp\left(-\frac{\mu_j^2}{t}\right)\\&+\left(\frac{\mathcal{A}(\Omega)}{4\pi}t
+\frac{l(\pa\Omega)}{8\sqrt{\pi}}t^{\frac{1}{2}}\right)\sum_{j=0}^{\infty}\exp\left(-\frac{m_j^2}{t}\right)\\
=&\frac{\mathcal{A}(\Omega)}{4\pi}t\sum_{\beta}\exp\left(-\frac{\sigma_{\beta}^2}{t}\right)-
\frac{l(\pa\Omega)}{8\sqrt{\pi}}t^{\frac{1}{2}}\sum_{\gamma}\exp\left(-\frac{\xi_{\gamma}^2}{t}\right).
\end{split}
\end{equation}The set of $\sigma_{\beta}^2$ contains $m_0^2$ with multiplicity two, $m_j^2, j\geq 1$, with multiplicity three, and $\mu_j^2, j\geq 1$ with multiplicity one. The set of $\xi_{\gamma}^2$ contains   $m_j^2, j\geq 1$,   and $\mu_j^2, j\geq 1$, each with multiplicity one.
On the other hand, when the cylinder is infinitely permeable, the set of $\tau_{\alpha}^2$ is given by (IP), which gives
\begin{equation}\label{eq2_8_2}\begin{split}
\sum_{\tau_{\alpha}}\exp\left(-\frac{\tau_{\alpha}^2}{t}\right)=&2\left(K_{\Omega,N}\left(t^{-1}\right)-1\right)\sum_{j=0}^{\infty}\!'
\exp\left(-\frac{m_j^2}{t}\right)+K_{\Omega,N}\left(t^{-1}\right)\sum_{j=1}^{\infty}\exp\left(-\frac{\mu_j^2}{t}\right)+ K_{\Omega,D}\left(t^{-1}\right) \sum_{j=0}^{\infty}
\exp\left(-\frac{m_j^2}{t}\right)\\&+\sum_{j=1}^{\infty}
\exp\left(-\frac{m_j^2}{t}\right)\\
\sim&\frac{\mathcal{A}(\Omega)}{4\pi}t\sum_{\beta}\exp\left(-\frac{\sigma_{\beta}^2}{t}\right)+
\frac{l(\pa\Omega)}{8\sqrt{\pi}}t^{\frac{1}{2}}\sum_{\gamma}\exp\left(-\frac{\xi_{\gamma}^2}{t}\right).
\end{split}\end{equation}Notice that the leading terms of \eqref{eq2_8_1} and \eqref{eq2_8_2} are the same and is given by
\begin{equation}\label{eq2_8_3}
\frac{\mathcal{A}(\Omega)}{4\pi}t\sum_{\beta}\exp\left(-\frac{\sigma_{\beta}^2}{t}\right).\end{equation}But the subleading terms
\begin{equation}\label{eq2_9_1}
\mp\frac{l(\pa\Omega)}{8\sqrt{\pi}}t^{\frac{1}{2}}\sum_{\gamma}\exp\left(-\frac{\xi_{\gamma}^2}{t}\right)
\end{equation}
have opposite signs.
Substituting \eqref{eq2_8_3} into \eqref{eq1_27_1}, we find that when $a\ll R$, the leading term of the Casimir force is given by
\begin{equation}\label{eq1_27_3}
\begin{split}
\mathcal{F}_0 (a)= & \frac{T\mathcal{A}(\Omega)}{8\pi^{\frac{3}{2}}}\frac{\pa}{\pa a} \left\{\frac{\sqrt{\pi}}{2}\frac{2+h}{a^2}\sum_{k=1}^{\infty}\frac{e^{2\pi i k\chi}}{k^3}
+2a \sum_{k=1}^{\infty}\sum_{\sigma_{\beta,p}\neq 0}e^{2\pi i k\chi}\left(\frac{\sigma_{\beta,p}}{ka}\right)^{\frac{3}{2}}K_{\frac{3}{2}}\left(2ka\sigma_{\beta,p}\right)\right\}\\
=&-\frac{T\mathcal{A}(\Omega)}{8\pi}\frac{2+h}{a^3}\sum_{k=1}^{\infty}\frac{e^{2\pi i k\chi}}{k^3}-\frac{T\mathcal{A}(\Omega)}{8\pi}
\sum_{k=1}^{\infty}\sum_{\sigma_{\beta,p}\neq 0}e^{2\pi i k\chi}\left(\frac{2\sigma_{\beta,p}^2}{ka}+\frac{2\sigma_{\beta,p}}{k^2a^2}+\frac{1}{k^3a^3}\right)e^{-2ka\sigma_{\beta,p}},
 \end{split}
\end{equation}where $\sigma_{\beta,p}^2=\sigma_{\beta}^2+[2\pi p T]^2$.  The first term in \eqref{eq1_27_3} comes from those terms with $\sigma_{\beta}^2=0$ and $p=0$. The sum
$$\sum_{k=1}^{\infty}\frac{e^{2\pi i k\chi}}{k^s}$$ is equal to $\zeta_R(s)$ if $\chi=1$ and is equal to $\displaystyle (2^{1-s}-1)\zeta_R(s)$ if $\chi=1/2$. \eqref{eq1_27_3} gives the leading behavior of the Casimir force when $a\ll R$ and $aT\gg 1$. If $aT\ll 1$, substituting \eqref{eq2_8_1} into \eqref{eq1_27_2} gives
\begin{equation} \label{eq2_8_4}
\begin{split}
\mathcal{F}_0 (a)=  &\frac{\mathcal{A}(\Omega)}{16\pi^2}\frac{\pa}{\pa a}\left\{(2+h)\left(\frac{1}{a^3}\sum_{k=1}^{\infty}\frac{e^{2\pi i k\chi}}{k^4}-\frac{8 \pi^4T^4a}{45}\right)
+2a\sum_{k=1}^{\infty}\sum_{\sigma_{\beta}\neq 0}e^{2\pi i k\chi}\left(\frac{\sigma_{\beta}}{ka}\right)^2K_2(2ka\sigma_{\beta})-8aT^2\sum_{p=1}^{\infty}
\sum_{\sigma_{\beta}\neq 0}\left(\frac{\sigma_{\beta}}{p}\right)^2\right.\\&\left.\times K_2\left(\frac{p\sigma_{\beta}}{T}\right)+2^{\frac{7}{2}}\sqrt{\pi}T^{\frac{3}{2}}
\sum_{p=1}^{\infty}\sum_{k=0}^{\infty}\sum_{\sigma_{\beta}}p^{-\frac{3}{2}}\left(\left[\frac{\pi(k+\chi)}{a}\right]^2+\sigma_{\beta}^2\right)^{\frac{3}{4}}
K_{\frac{3}{2}}\left(\frac{p}{T}\sqrt{\left[\frac{\pi(k+\chi)}{a}\right]^2+\sigma_{\beta}^2}\right)\right\}\\
=&-(2+h)\mathcal{A}(\Omega)\left(\frac{3}{16\pi^2a^4}\sum_{k=1}^{\infty}\frac{e^{2\pi i k\chi}}{k^4}+\frac{  \pi^2T^4 }{90}\right)
-\frac{3\mathcal{A}(\Omega)}{8\pi^2}\sum_{k=1}^{\infty}\sum_{\sigma_{\beta}\neq 0}e^{2\pi i k\chi}\left(\frac{\sigma_{\beta}}{ka}\right)^2K_2(2ka\sigma_{\beta})
\\&-\frac{ \mathcal{A}(\Omega)}{4\pi^2}\sum_{k=1}^{\infty}\sum_{\sigma_{\beta}\neq 0} e^{2\pi i k\chi} \frac{\sigma_{\beta}^3}{ka} K_1(2ka\sigma_{\beta})
 -\frac{\mathcal{A}(\Omega)T^2}{2\pi^2}\sum_{p=1}^{\infty}
\sum_{\sigma_{\beta}\neq 0}\left(\frac{\sigma_{\beta}}{p}\right)^2 K_2\left(\frac{p\sigma_{\beta}}{T}\right)
\\&+\frac{\pi \mathcal{A}(\Omega)T}{2a^3}\sum_{p=1}^{\infty}\sum_{k=0}^{\infty}\sum_{\sigma_{\beta}}\frac{(k+\chi)^2}{p}
\exp\left(-\frac{p}{T}\sqrt{\left[\frac{\pi(k+\chi)}{a}\right]^2+\sigma_{\beta}^2}\right).
\end{split}
\end{equation}

For the subleading term $\mathcal{F}_1(a)$, substitute  \eqref{eq2_9_1} into \eqref{eq1_27_1} and \eqref{eq1_27_2} respectively. Similar computations show that
\begin{equation*}
\begin{split}
\mathcal{F}_1(a)=\pm \frac{Tl(\pa\Omega)}{16\pi}\left\{\frac{h}{a^2}\sum_{k=1}^{\infty}\frac{e^{2\pi i k\chi}}{k^2}+2\sum_{k=1}^{\infty}\sum_{\xi_{\gamma,p}\neq 0}e^{2\pi i k\chi}\frac{\xi_{\gamma,p}}{ka}K_1(2ka\xi_{\gamma,p})+4\sum_{k=1}^{\infty}\sum_{\xi_{\gamma,p}\neq 0}e^{2\pi i k\chi} \xi_{\gamma,p}^2K_0(2ka\xi_{\gamma,p})\right\},
\end{split}
\end{equation*}with $\xi_{\gamma,p}^2=\xi_{\gamma}^2+[2\pi pT]^2$, or
\begin{equation*}
\begin{split}
\mathcal{F}_1(a)=&\pm l(\pa\Omega)\left\{ \frac{h}{32\pi} \left(\frac{1}{a^3}\sum_{k=1}^{\infty}\frac{e^{2\pi i k\chi}}{k^3}+4T^3\zeta_R(3)\right)+\frac{1}{32\pi}
\sum_{k=1}^{\infty}\sum_{\xi_{\gamma}\neq 0}e^{2\pi i k\chi}e^{-2ka\xi_{\gamma}}\left(\frac{2\xi_{\gamma}^2}{ka}+\frac{2\xi_{\gamma}}{k^2a^2}+\frac{1}{k^3a^3}\right)
\right.\\&\left.+\frac{ T^2}{8\pi}\sum_{p=1}^{\infty}\sum_{\xi_{\gamma}\neq 0}e^{-\frac{p\xi_{\gamma}}{T}}\left(\frac{\xi_{\gamma}}{p^2}+\frac{T}{p^3}\right)-\frac{ \pi}{4a^3}\sum_{k=0}^{\infty}\sum_{p=1}^{\infty}\sum_{\xi_{\gamma}}(k+\chi)^2K_0\left(\frac{p}{T}\sqrt{\left[\frac{\pi(k+\chi)}{a}\right]^2+\xi_{\gamma}^2}\right)\right\},
\end{split}
\end{equation*}
where the plus sign is for perfectly conducting cylinder, and the minus sign is for infinitely permeable cylinder.

To find the asymptotic behavior when $a\ll r, R$, we use the fact that
\begin{equation*}
\begin{split}
\sum_{\beta}\exp\left(-\frac{\sigma_{\beta}^2}{t}\right)=&2+3\sum_{j=1}^{\infty}\exp\left(-\frac{m_j^2}{t}\right)+\sum_{j=1}^{\infty}\exp\left(-\frac{\mu_j^2}{t}\right)
=2K_{\mathcal{N},0}\left(t^{-1}\right)+K_{\mathcal{N},1}\left(t^{-1}\right)\sim (n+2)\frac{\mathcal{V}(\mathcal{N}) }{(4\pi)^{\frac{n}{2}}}t^{\frac{n}{2}}\\
\sum_{\gamma}\exp\left(-\frac{\xi_{\gamma}^2}{t}\right)=&\sum_{j=1}^{\infty}\exp\left(-\frac{m_j^2}{t}\right)+\sum_{j=1}^{\infty}\exp\left(-\frac{\mu_j^2}{t}\right)
=K_{\mathcal{N},1}\left(t^{-1}\right)\sim n\frac{\mathcal{V}(\mathcal{N}) }{(4\pi)^{\frac{n}{2}}}t^{\frac{n}{2}}.
\end{split}
\end{equation*}Therefore,
\begin{equation}\label{eq2_10_1}
\begin{split}
\sum_{\alpha}\exp\left(-\frac{\tau_{\alpha}^2}{t}\right)\sim & (n+2)\frac{\mathcal{V}(\mathcal{S})  }{(4\pi)^{\frac{n+2}{2}}}t^{\frac{n+2}{2}}
\mp n\frac{\mathcal{V}(\pa\mathcal{S})  }{4(4\pi)^{\frac{n+1}{2}}}t^{\frac{n+1}{2}},
\end{split}
\end{equation}where the minus sign is for perfectly conducting cylinder,   the plus sign is for infinitely permeable cylinder; $\mathcal{S}=\Omega\times\mathcal{N}$ is the cross section of the cylinder, and $\mathcal{V}(\mathcal{S})=\mathcal{A}(\Omega)\times\mathcal{V}(\mathcal{N})$ is its volume.
Substituting \eqref{eq2_10_1} into \eqref{eq1_27_1} and \eqref{eq1_27_2}, we find that the first two leading terms of the Casimir force when $a\ll r,R$ are given by
\begin{equation*}
\begin{split}
\mathfrak{F}_0(a) = (n+2) \mathcal{V}(\mathcal{S})&\left\{-T\frac{(n+2)\Gamma\left(\frac{n+3}{2}\right)}{(4\pi)^{\frac{n+3}{2}}a^{n+3}}\sum_{k=1}^{\infty}
\frac{e^{2\pi i k\chi}}{k^{n+3}}-(n+2)\frac{T^{\frac{n+5}{2}}}{2^{\frac{n-1}{2}}}\sum_{k=1}^{\infty}\sum_{p=1}^{\infty}e^{2\pi i k\chi}\left(\frac{p}{ka}\right)^{\frac{n+3}{2}}K_{\frac{n+3}{2}}(4\pi kp aT)\right.\\
&\left.-\frac{\pi T^{\frac{n+7}{2}}}{2^{\frac{n-5}{2}}}\sum_{k=1}^{\infty}\sum_{p=1}^{\infty}e^{2\pi i k\chi} \frac{p^{\frac{n+5}{2}}}{(ka)^{\frac{n+1}{2}}}K_{\frac{n+1}{2}}(4\pi kp aT)\right\},
\\
\mathfrak{F}_1(a)=\mp\frac{n\mathcal{V}(\pa\mathcal{S})}{4}&\left\{-T\frac{(n+1)\Gamma\left(\frac{n+2}{2}\right)}{(4\pi)^{\frac{n+2}{2}}a^{n+2}}\sum_{k=1}^{\infty}
\frac{e^{2\pi i k\chi}}{k^{n+2}}-(n+1)\frac{T^{\frac{n+4}{2}}}{2^{\frac{n-2}{2}}}\sum_{k=1}^{\infty}\sum_{p=1}^{\infty}e^{2\pi i k\chi}\left(\frac{p}{ka}\right)^{\frac{n+2}{2}}K_{\frac{n+2}{2}}(4\pi kp aT)\right.\\
&\left.-\frac{\pi T^{\frac{n+6}{2}}}{2^{\frac{n-6}{2}}}\sum_{k=1}^{\infty}\sum_{p=1}^{\infty}e^{2\pi i k\chi} \frac{p^{\frac{n+4}{2}}}{(ka)^{\frac{n}{2}}}K_{\frac{n}{2}}(4\pi kp aT)\right\},
\end{split}
\end{equation*}
or
\begin{equation*}
\begin{split}
\mathfrak{F}_0(a) = (n+2) \mathcal{V}(\mathcal{S})&\left\{- \frac{(n+3)\Gamma\left(\frac{n+4}{2}\right)}{(4\pi)^{\frac{n+4}{2}}a^{n+4}}\sum_{k=1}^{\infty}
\frac{e^{2\pi i k\chi}}{k^{n+4}}-\frac{\Gamma\left(\frac{n+4}{2}\right)}{\pi^{\frac{n+4}{2}}}\zeta_R(n+4)T^{n+4}\right.\\&\left.+\frac{\pi T^{\frac{n+1}{2}}}{2^{\frac{n+1}{2}}a^{\frac{n+7}{2}}}\sum_{k=0}^{\infty}\sum_{p=1}^{\infty}\frac{(k+\chi)^{\frac{n+5}{2}}}{p^{\frac{n+1}{2}}}K_{\frac{n+1}{2}}
\left(\frac{\pi p(k+\chi)}{aT}\right)\right\},
\\
\mathfrak{F}_1(a) =\mp\frac{ n \mathcal{V}(\pa\mathcal{S})}{4}&\left\{- \frac{(n+2)\Gamma\left(\frac{n+3}{2}\right)}{(4\pi)^{\frac{n+3}{2}}a^{n+3}}\sum_{k=1}^{\infty}
\frac{e^{2\pi i k\chi}}{k^{n+3}}-\frac{\Gamma\left(\frac{n+3}{2}\right)}{\pi^{\frac{n+3}{2}}}\zeta_R(n+3)T^{n+3}\right.\\&\left.+\frac{\pi T^{\frac{n}{2}}}{2^{\frac{n}{2}}a^{\frac{n+6}{2}}}\sum_{k=0}^{\infty}\sum_{p=1}^{\infty}\frac{(k+\chi)^{\frac{n+4}{2}}}{p^{\frac{n}{2}}}K_{\frac{n}{2}}
\left(\frac{\pi p(k+\chi)}{aT}\right)\right\}.
\end{split}
\end{equation*}

\begin{acknowledgments}
We have benefited from discussions with K. Kirsten. This project is funded by Ministry of Higher Education of Malaysia   under FRGS grant.
\end{acknowledgments}

\end{document}